\documentclass[onecolumn,nofootinbib,preprintnumbers, sort&compress, floatfix,a4paper,superscriptaddress,9pt]{revtex4-1}
\usepackage[framemethod=TikZ]{mdframed}
\usepackage[most]{tcolorbox}
\usepackage{longtable}
\usepackage{mathcomp}
\usepackage{pifont}
\usepackage{xcolor}
\usepackage{amsthm}
\usepackage{pgf}
\usepackage{pgfpages}
\usepackage{amsmath}
\usepackage{amsfonts}
\usepackage{amssymb}
\usepackage{graphicx}
\usepackage{bm}
\usepackage{hyperref}
\usepackage{lipsum}
\usepackage{array}
\usepackage{booktabs}
\usepackage{tikz}
\usepackage{titlesec}
\usepackage{hyperref}
\hypersetup{bookmarksnumbered}
\def\be{\begin{equation}}
\def\ee{\end{equation}}
\def\bi{\bibitem}
\usepackage{amsmath}
\usepackage{empheq}
\usepackage[most]{tcolorbox}
\newtcbox{\mymath}[1][]{%
    nobeforeafter, math upper, tcbox raise base,
    enhanced, colframe=blue!30!black,
    colback=blue!30, boxrule=1pt,
    #1}
\usepackage{mdframed}
\usepackage{lipsum}
\usepackage{PGF}
\usepackage{pgfkeys}
\usepackage{pgffor}
\usepackage{pgfcalendar}
\usepackage{pgfpages}
\usepackage{tikz}
\usetikzlibrary{calc}
\usepackage{array}
\usepackage{booktabs}
\usepackage{multirow}
\usepackage{latexsym}
\usepackage[english]{babel}
\allowdisplaybreaks
\vbox{\hbox{1000}}
\hypersetup{colorlinks=true,linkcolor=blue!100,citecolor=red!100}
\newcounter{theo}[section] 
\renewcommand{\thetheo}{\arabic{section}.\arabic{theo}}
\newenvironment{theo}[2][]{%
\refstepcounter{theo}%
\ifstrempty{#1}%
{\mdfsetup{%
frametitle={%
\tikz[baseline=(current bounding box.east),outer sep=0pt]
\node[anchor=east,rectangle,fill=blue!20]
{\strut Theorem~\thetheo};}}
}%
{\mdfsetup{%
frametitle={%
\tikz[baseline=(current bounding box.east),outer sep=0pt]
\node[anchor=east,rectangle,fill=blue!20]
{\begin{minipage}{0.99\linewidth}Theorem~\thetheo:~#1\end{minipage}};}}%
}%
\mdfsetup{innertopmargin=10pt,linecolor=blue!20,%
linewidth=2pt,topline=true,%
frametitleaboveskip=\dimexpr-\ht\strutbox\relax
}
\begin{mdframed}[]\relax%
\label{#2}}{\end{mdframed}}

\newcounter{prf}[section]
\renewcommand{\theprf}{\arabic{section}.\arabic{prf}}
\newenvironment{prf}[2][]{%
\refstepcounter{prf}%
\ifstrempty{#1}%
{\mdfsetup{%
frametitle={%
\tikz[baseline=(current bounding box.east),outer sep=0pt]
\node[anchor=east,rectangle,fill=red!20]
{\strut Proof~\theprf};}}
}%
{\mdfsetup{%
frametitle={%
\tikz[baseline=(current bounding box.east),outer sep=0pt]
\node[anchor=east,rectangle,fill=red!20]
{\strut Proof~\thetheo:~#1};}}%
}%
\mdfsetup{innertopmargin=10pt,linecolor=red!20,%
linewidth=2pt,topline=true,%
frametitleaboveskip=\dimexpr-\ht\strutbox\relax
}
\begin{mdframed}[]\relax%
\label{#2}}{\end{mdframed}}
\newcounter{lem}[section]
\renewcommand{\thelem}{\arabic{section}.\arabic{lem}}

\titleformat{\section}
{\color{black}\normalfont\Large\bfseries}
{\color{black}\thesection}{1em}{}
\renewcommand\thesection{\arabic{section}}

\newcommand*{\hrmybox}[2]{\colorbox[rgb]{0.70,0.70,1.00}{\parbox{.99\linewidth}{#2}}}
\newcommand*{\rmybox}[2]{\colorbox[rgb]{0.91,0.45,0.45}{\parbox{.99\linewidth}{#2}}}

\newcommand*{\pqrmybox}[2]{\colorbox[rgb]{0.00,0.98,0.25}{\parbox{.99\linewidth}{#2}}}
\newcommand*{\goldmybox}[2]{\colorbox[rgb]{0.76,0.98,0.25}{\parbox{.99\linewidth}{#2}}}

\allowdisplaybreaks
\usepackage{booktabs}

\begin{document}
\title{\textcolor[rgb]{0.00,0.40,0.80}{\textsf{\huge\boldmath A new approach to the analysis of the reconstruction methods, phase space, and exact solutions of the alternative theories of gravity.\\}}}
\author{‪Behzad Tajahmad}
\email{behzadtajahmad@yahoo.com}
\affiliation{Faculty of Physics, University of Tabriz, Tabriz, Iran}
\affiliation{Research Institute for Astronomy and Astrophysics of Maragha (RIAAM)-Maragha, Iran, P.O. Box: 55134-441\\}
\begin{center}
\begin{abstract}
\begin{tcolorbox}[breakable,colback=white,
colframe=cyan,width=\dimexpr\textwidth+0mm\relax,enlarge left by=-17mm,enlarge right by=-6mm ]
\text{ }\\
\large{\textbf{\textsf{Abstract:}}
A new approach, ``$\mathfrak{B}\text{-Function}$ Method'', to the analysis of the reconstruction methods, phase space, and exact solutions of the alternative theories of gravity is suggested.
The $\mathfrak{B}\text{-function}$ method which is constructed by suggesting eight new model-independent physical theorems, extracts the physically admissible ranges of the evolution of the parameters (i.e. the variables and constant parameters of action and those which emerged by differential equations as the constants of integration) and helps one for quick data analysis. The most significant feature and beauty of the $\mathfrak{B}\text{-function}$ method are that all conditions leading to physically admissible domains, can be expressed only by one new function, $\mathfrak{B}\text{-Function}$, which is the Lie derivative of $\ln(\mathcal{F}(x))$ along a vector field, $\mathbf{X}=x+c$, on a singleton $Q=\{x\}$.\\} \end{tcolorbox}
\end{abstract}

\end{center}
\maketitle
\hrule \hrule \hrule
\textbf{\textcolor[rgb]{0.00,0.00,0.00}{\tableofcontents}}
\text{ }
\hrule \hrule \hrule
\noindent\hrmybox{}{\section{Introduction\label{sec:intro}}}\vspace{5mm}

Various astronomical and cosmological observations of the last decade, including CMB studies \cite{CMB}, supernova type Ia \cite{sup1,sup2}, baryon acoustic oscillations \cite{baryon}, weak lensing \cite{lens}, and large-scale structure \cite{LSC} have provided a picture of the universe with accelerating expansion. As we know, the explanation of the essence and mechanism of the acceleration of our universe is one of the great and major challenges for physicists.

The accelerating expansion of the universe is driven by mysterious energy with negative pressure known as Dark Energy (DE) \cite{DE1}. Two important problems like `fine-tuning' and `cosmic coincidence' are related to dark energy. It is thought that the most probable solution to DE is the Einstein's cosmological constant \cite{Sol-DE}, but it cannot resolve
the two problems mentioned above. Hence, other theoretical models by considering the dynamic nature of dark energy like quintessence \cite{quin1, quin2, quin3}, phantom field \cite{phantom1, phantom2, phantom3, phantom4, phantom5, phantom6}, tachyon field \cite{tachyon}, quintom \cite{quintom1, quintom2, quintom3, quintom4}, and the interacting DE models like Chaplygin gas \cite{Chap}, holographic models \cite{holo1,holo2}, braneworld models \cite{brane}, and etcetera, have been proposed to interpret accelerating universe. Another interesting set of proposals to DE puzzle which was proposed after the failure of general relativity (GR), is the `modified gravity' \cite{MG1, MG2}, making the action of the theory dependent upon a function of the curvature scalar $R$; in a certain limit of the parameters, the theory reduces to general relativity. Recently, various gravitational modification theories like $F(R)\text{-gravity}$, $F(T)\text{-gravity}$, $F(T)\text{-gravity}$ with an unusual term \cite{beh-intervention}, and scalar-tensor theories, have been lionized \cite{lio}. This new set of gravity theories passes several solar system and astrophysical tests successfully \cite{solar1,solar2,solar3}.

Scalar fields play an important role in unified theories of interactions and also in inflationary scenarios in cosmology. Indeed, a rich variety of dark energy and inflationary models may be accommodated phenomenologically by scalar fields in which the inflations produce the initial acceleration. Many cosmological models and modified gravity theories with scalar fields involve some scalar functions especially scalar-field functions which cannot be derived easily from the basic theory. The choice of the unknown functions, somewhat arbitrary, has given rise to the objection of fine-tuning, the very problem whose solutions have been set out through inflationary theories. Therefore, it is favorable to have a path to extract the potential or at least some criteria for admissible potentials. In this respect, two well-known approaches namely Noether symmetry approach and the reconstruction technique have been suggested.\\
The Noether symmetry approach was applied by many authors and lead to graceful results (for example, see Refs. \cite{g51,no2,no3,no4,no5,no6,no7,no8,no9,no10,no11,no12,no13,no14,no15,no16,
no17,Y1no17,Y2no17,Y3no17,no18,no19,no20,no21,no22,no23,no24,no25,no26,no27,
no28,no29,no30,no31,no32,no33,no34,no35,no36,no37,no38}. In the Noether process, besides finding the form of unknown functions, the conserved currents of the system is also obtained. However, some hidden conserved currents may not be obtained by the Noether symmetry approach \cite{g17, g31}. It is worthwhile mentioning that beside the Noether symmetry approach, another approach, the Beyond Noether Symmetry approach (B.N.S. approach), has recently been presented \cite{g51}. The B.N.S. approach may carry more conserved currents than the Noether symmetry approach. Furthermore, sometimes the Noether symmetry approach fails to achieve the purpose. In such cases, using the B.N.S. approach would be so helpful.\\
Like the Noether symmetry approach, the technique of the reconstruction of the potentials of scalar fields has also been taken into account. This technique enables one to find the form of the scalar field potential as well as the scalar field for a particular value of the Hubble parameter in terms of scale factor or cosmic time, or particularly the redshift. For instance, in ref. \cite{ref1}, the Hubble parameter arisen from `Barotropic fluid', `Cosmological constant', `Chaplygin gas', and `Modified Chaplygin gas' have been investigated. Furthermore, the reconstruction of scalar-field dark energy models from observations has attracted the attention of researchers for a long time \cite{367, 368, 369, 370, 371}. In fact, one can reconstruct the potential and the equation of state of the field by parameterizing the Hubble parameter $H$ in terms of the redshift $z$ \cite{372}. The Hubble rate $H(z)$ is determined by the luminosity distance $d_L(z)$ by using the relation
\begin{align*}
H(z)=\left[\frac{\mathrm{d}}{\mathrm{d}z}\left(\frac{d_{L}(z)}{1+z} \right) \right]^{-1}.
\end{align*}
If we measure the luminosity distance observationally, we may determine the expansion rate of the universe. This method was generalized to scalar-tensor theories \cite{354, 355, 373.1, 373.2}, $F(R)$ gravity \cite{374,venus1r,venus2r,venus3r}, and also a dark-energy fluid with viscosity terms \cite{375}. The reconstruction of scalar theory (actually potentials) for different evolutions was given in \cite{venus1p,venus2p,ref1}. Furthermore, a bottom-up \(F(R)\) gravity reconstruction technique has recently been introduced by S.D. Odintsov and V.K. Oikonomou \cite{1368}.\\

\noindent\hrmybox{}{\section{$\mathfrak{B}\text{-Function}$ Method\label{The A.R.M.}}}\vspace{5mm}

In the literature, the analysis way of the reconstruction method has not been presented. In this section, we would like to suggest a new approach to the analysis of the reconstruction method. We named it the ``$\mathfrak{B}\text{-Function}$ Method''. The $\mathfrak{B}\text{-function}$ method is a model-independent analysis. Furthermore, the $\mathfrak{B}\text{-function}$ method may also be applicable to other subjects of alternative theories of gravity as well (for example, dynamical systems (phase space), the analysis of the exact solutions of alternative theories, and etcetera). In a nutshell, the $\mathfrak{B}\text{-function}$ method is a new analysis method to alternative theories of gravity. Now, let us explain this approach.\\
While I was dealing with the reconstruction method, I found this approach, hence I would like to explain the approach by starting from the reconstruction perspective.\\
Roughly speaking, the systems of alternative theories of gravity are so-called dynamical systems, hence the elements of configuration space of the problem (e.g. the scale factor ($a$) and the scalar field ($\varphi$)) are naturally dependent on time. In the extended reconstruction method proposed in ref. \cite{ref1}, this dependence on the basic variable (i.e. time) is substituted by the scale factor and it causes that the treatments of almost all parameters versus time be hidden. Hence, in order to analyze the parameters, knowing the treatments of the elements of the configuration space versus time is necessary. The behavior of the scale factor is obvious, because of the observational data, namely the status of the expansion and the phase of the universe, but, the scalar field is unclear. It may be easily understood that as long as its treatment versus time is unclear, the behaviors of almost all parameters and also their physical domains would be obscure. Indeed, in the reconstruction method, the behaviors of almost all parameters are affected by the treatment of the scalar field. Hence, in order to analyze the parameters, we have to impose a treatment for the scalar field, and then obtain the behaviors of other parameters. Note that this may also be regarded as a `reconstruction'. It may be demonstrated that the nature of the scalar field at each very short time interval, generates two separate classes for the reconstruction analysis: \emph{i}- Class-1 (Table (\ref{RMA})): Based on the strictly \textbf{D}ecreasing \textbf{S}calar \textbf{F}ield with time (D.S.F.); and \emph{ii}- Class-2 (Table (\ref{RMA2})): Based on the strictly \textbf{I}ncreasing \textbf{S}calar \textbf{F}ield with time (I.S.F.). However, the scalar field might be a mixture of both classes in throughout of its evolution range (i.e. in some eras be dependent on `Class-1' and in other eras be dependent on `Class-2'). In such hybrid cases, we can separate it into monotonic parts. Nonetheless, since `we' reconstruct the potentials, hence `we' should impose that which class is our objective. Clearly, choosing a hybrid case is cumbersome. Furthermore, in many papers, if the `exact' data analysis is done, then it will appear that the most of the scalar fields are of decreasing nature. Altogether, it is better we restrict ourselves into monotonic cases (i.e. `Class-1' or `Class-2'), for simplicity. In other words, indeed, it is more convenient that we reconstruct the behaviors of all parameters (both variables and constants) based on one of the treatments of the strictly monotonic increasing/decreasing scalar fields in the whole of the time interval of interest. Note that it may be done when the form of the scalar field in terms of the time is not accessible, otherwise the type of the class shall appear spontaneously.\\
It must be noted that there are other reconstruction approaches in which the redshift $z$ or $N=\ln(a)$ is demanded to be substituted instead of time $t$. For this problem, we make the $\mathfrak{B}\text{-function}$ method in a manner that it covers such other approaches as well.\\
Each class of the $\mathfrak{B}\text{-function}$ method is built on four new theorems. The theorems and their proofs are presented in the ``Supplement: Proof of the $\mathfrak{B}\text{-Function}$ Method; Sect. (\ref{Proofss})''. The physical treatments of all the parameters may be described by defining a new `dimensionless' function, $\mathfrak{B}=\mathfrak{B}[x,c;\mathcal{F}]$, as\footnote{We assume that $\ln(\mathcal{F}(x))$ is at least $C^{1}$ at every point in $x\text{-interval}$ of interest.}
\begin{empheq}[box={\mymath[colback=blue!15,drop lifted shadow]}]{equation}\label{lie12}
\mathfrak{B}[x,c;\mathcal{F}(x)] \stackrel{\text{def.}}{=\joinrel=} \pounds_{\mathbf{X}} \ln(\mathcal{F}(x)),
\end{empheq}
where $c$ is an arbitrary constant, $\mathcal{F}=\mathcal{F}(x)$ is a function of the first argument of the $\mathfrak{B}\text{-function}$ (i.e. $x$), and $\pounds_{\mathbf{X}} \ln(\mathcal{F}(x))$  is the Lie derivative of $\ln(\mathcal{F}(x))$ along a vector field $\mathbf{X}$ on a singleton $Q=\{x\}$. We define this vector field as
\begin{align}\label{lie11}
\mathbf{X}\equiv (x+c) \frac{\mathrm{d}}{\mathrm{d}x}.
\end{align}
Note that the variable $x$ and the constant $c$ have the same dimensions. We named it $\mathfrak{B}\text{-function}$ since it determines the \textbf{B}oundaries of the parameters. The most significant feature of the $\mathfrak{B}\text{-function}$ method is that all the physical conditions are expressed only by this function.\\
The vector field (\ref{lie11}) signifies an infinitesimal point transformation. Any smooth and invertible point transformation of the generalized coordinates $q^{i} \to Q^{i}(\mathbf{q})$ induces a transformation of the generalized velocities\footnote{It has been assumed that the Jacobian matrix $J= \| \partial Q^{i}/ \partial q^{j} \|$ of the coordinate transformation does not vanish. Note that $J \neq 0$ implies that the Jacobian of the induced transformation is also nonzero. However, in general, this transformation is local since $J \neq 0$ is satisfied only in the neighborhood of a given point, not on the entire space. Nonetheless, if the transformation is extended to the maximal sub-manifold on which $J \neq 0$, then problems may arise for the whole manifold due to the possibility of different topologies.}
\begin{align}\label{doz12}
\dot{q}^{i} \to \dot{Q}^{i}(\mathbf{q})=\frac{\partial Q^{i}}{\partial q^{j}}\dot{q}^{j}.
\end{align}
where a dot denotes derivative with respect to the affine parameter $\lambda$ which usually corresponds to the cosmic time $t$. According to (\ref{lie11}), this induced transformation (\ref{doz12}) is represented by
\begin{align}\label{lie14}
\mathbf{X}^{(c)}=(x+c) \frac{\mathrm{d}}{\mathrm{d}x}
+\left(\frac{\mathrm{d}(x+c)}{\mathrm{d}\lambda}\right)
\frac{\mathrm{d}}{\mathrm{d}\dot{x}}.
\end{align}
in which the vector field $\mathbf{X}^{(c)}$ is called the complete lift of $\mathbf{X}$. Therefore, it is not hard to show that the Lie derivative of $\ln(\mathcal{F}(x))$ along the complete lift of $\mathbf{X}$ is also equal to $\mathfrak{B}[x,c;\mathcal{F}(x)]$, viz.,
\begin{align}\label{lie13}
\mathfrak{B}[x,c;\mathcal{F}(x)]=\pounds_{\mathbf{X}^{(c)}} \ln(\mathcal{F}(x)).
\end{align}

Pursuant to the vector field (\ref{lie11}), the definition of $\mathfrak{B}\text{-function}$ in (\ref{lie12}) may be recast in the form
\begin{empheq}[box={\mymath[colback=blue!15,drop lifted shadow]}]{equation}\label{deltafunction}
\mathfrak{B}[x,c;\mathcal{F}(x)] \stackrel{\text{def.}}{=\joinrel=} (x+c) \; \frac{\mathrm{d} \ln(\mathcal{F})}{\mathrm{d}x}
\end{empheq}
which can be regarded as the effect of a special form of the Cauchy-Euler operator, which is a differential operator of the form
\begin{equation}\label{CauchyEuler}
\mathbb{P}(x)\frac{\mathrm{d}}{\mathrm{d}x}
\end{equation}
where $\mathbb{P}(x)$, in general, is a polynomial (i.e. $\mathbb{P}(x)=\sum_{k=0}^{n}\lambda_{k}\;x^{k}$), on the logarithmic function of $\mathcal{F}(x)$.

In the case that both $x$ and $c$ are dimensionless, it is convenient to define another new `dimensionless' function as
\begin{equation}\label{BStar}
\mathfrak{B}^{\star}[x,c;\mathcal{F}(x)] \stackrel{\text{def.}}{=\joinrel=} \frac{\mathfrak{B}[x,c;\mathcal{F}(x)]}{\mathbf{X} \; x}
\stackrel{\text{cf.(\ref{lie11})}}{=\joinrel=\joinrel=}
\frac{\mathfrak{B}[x,c;\mathcal{F}(x)]}{x+c}.
\end{equation}
However, $\mathfrak{B}^{\star}$ is not an independent function, but we defined it to shorten the writing of equations. For example, the terms like $y_{,N}/y$ where $N=\ln(a)$ (which is a dimensionless parameter) can be written as $\mathfrak{B}^{\star}[N,0;y(N)]$ instead of $(\mathfrak{B}[N,0;y(N)])/N$. The main context of the application of $\mathfrak{B}^{\star}$ is in the investigation of the phase space of the alternative theories of gravity (See `Example 2' in Sub-section (\ref{yaasin})).\\

\begin{figure*}
\centering
\includegraphics[width=500 px, height=500 px]{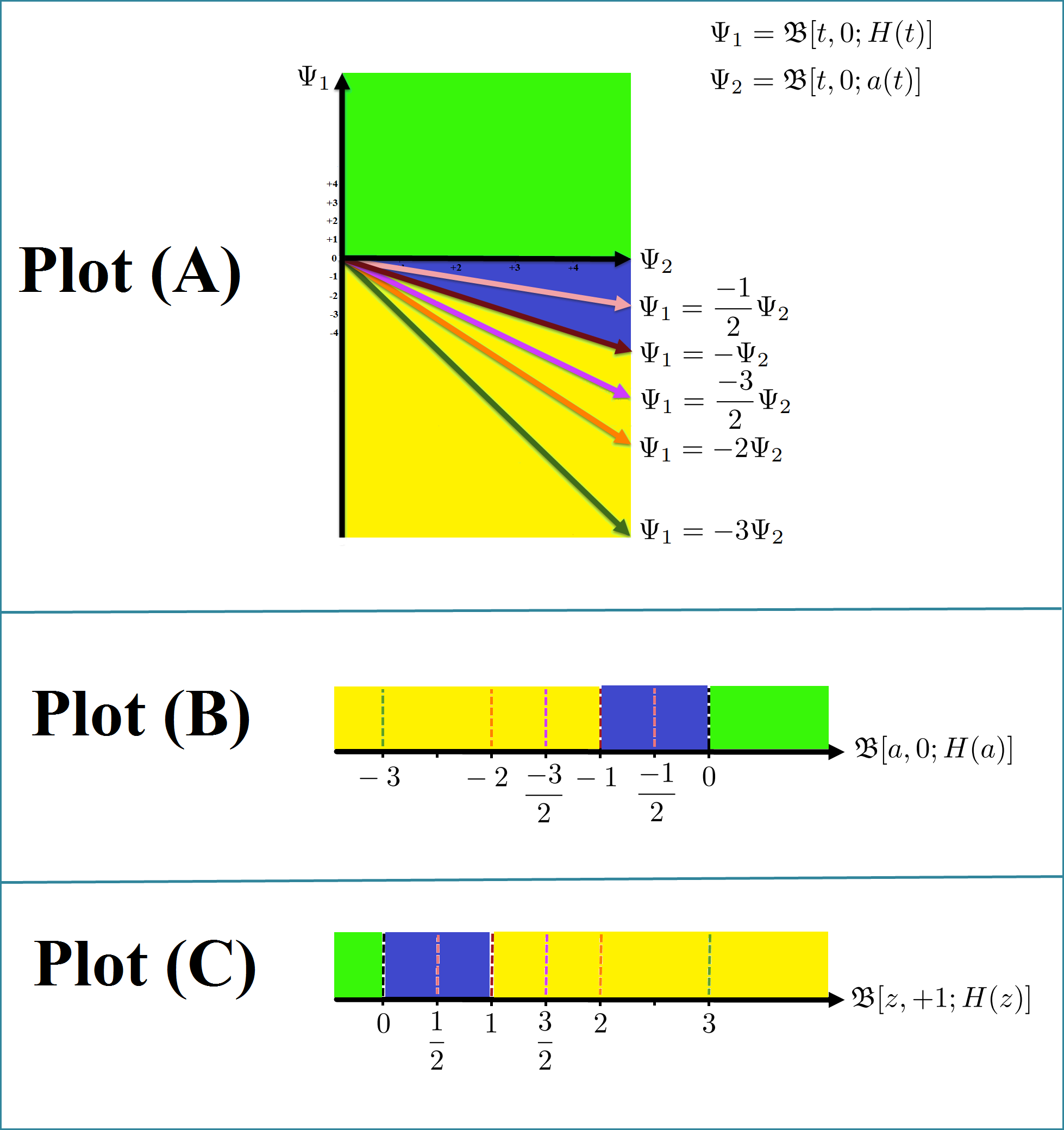}\\
\caption{In this figure, the different statuses of the universe are presented in three plots (`Plot (A)' demonstrates the states of the universe according to the amounts of the $\Psi_{1}=\mathfrak{B}[t,0;H(t)]$ versus the amounts of the $\Psi_{2}=\mathfrak{B}[t,0;a(t)]$; `Plot (B)' and `Plot (C)' indicate equivalent areas to `Plot (A)' in terms of $\mathfrak{B}[a,0;H(a)]$ and $\mathfrak{B}[z,+1;H(z)]$, respectively). The physical meanings of areas and lines are presented in the Table (\ref{yastable}).}\label{fig7}
\end{figure*}

\textcolor[rgb]{1.00,0.00,0.50}{\ding{228}}
\textcolor[rgb]{1.00,0.00,1.00}{\textbf{The prescription for applying the $\mathfrak{B}\text{-function}$ in reconstruction methods:}}\\
The first step of the procedure of the $\mathfrak{B}\text{-function}$ method in the context of reconstruction methods like ref. \cite{ref1} is that one must specify that he prefers which class. As already noticed, the type of class is given spontaneously when we know the scalar field in terms of the time (for example, when the analysis of the exact solutions is demanded), otherwise, a treatment should be imposed (for example, see `Example 1: Application in the analysis of the reconstruction method' in sub-section (\ref{example114})). According to both tables (\ref{RMA}) and (\ref{RMA2}), when one has $\varphi(t)$, then the sign of the $\mathfrak{B}[t,0;\varphi(t)]$ shall specify the type of class. Note that both classes are not correct at precisely the same time, because at a very short interval, a function can be monotonically increasing or decreasing versus its variable. The second step is that all the \textit{`applicable'}\footnote{I use the `applicable' word to mention that the tables are presented in a way that they are able to cover other reconstruction procedures and other applications as well, therefore, there are some cases in which we do not know some forms of functions, so their related conditions cannot be investigated, for example, in the first example of this paper, we do not have access to $\varphi(t)$.}\edef\thefndolors{\the\value{footnote}} conditions demonstrated in the relevant table of the chosen class and also the common table, namely table (\ref{yastable}) which is the descriptions of special areas and lines of the plots (A), (B), and (C) in the figure (\ref{fig7}), should be considered. Before considering the common table (\ref{yastable}), it is better we plot the feasible class-independent portraits like the presented plots in the figure (\ref{fig7}) and then compare our plots with the figure (\ref{fig7}), and then read their physical meaning from the table (\ref{yastable}).\\
In the anisotropic universes, for each direction, the conditions must be established separately\footnote{For example, for the Locally Rotationally Symmetric Bianchi type I background geometry, $\mathrm{d}s^2=\mathrm{d}t^2-A^2(t)\mathrm{d}x^2-B^2(t) \left[\mathrm{d}y^2+\mathrm{d}z^2\right]$, `$a$' can be $A$ and also $B$, and consequently, $H(a)$ can be $H_{1}(A)$ (The directional Hubble parameter along $x$ direction) and $H_{2}(B)$ (The directional Hubble parameter along $y$ and $z$ directions), respectively.}. And finally, for each spatial direction, the domains of constant parameters and also the evolution ranges of variable parameters should be chosen from the `common domains' of all applicable conditions. It is worth mentioning that because the scale factor of the different directions are generally independent of each other, hence the common domains obtained for each direction is only for that direction, in other words, in general (when there is no relation among the scale factors of spatial directions), common domains of parameters of, for example, $x\text{-direction}$ should not be mixed with common domains of $y$ and also $z$ directions.\\
Note that in both classes, there are two types of scalar functions, namely $i$. Decreasing scalar-field function with respect to time, and $ii$. Increasing scalar-field function with respect to time. This scalar-field function can be the scalar field potential $V(\varphi)$, and also $\varphi\text{-dependent}$ coupling functions with curvature/torsion/electromagnetic field and etcetera. In each class, these types of functions cannot simultaneously be correct (i.e. At each very short interval, the scalar function can be Type-I or Type-II). Therefore, at each very short time interval, the behavior of each scalar-field function in each class generates two sub-classes. Note that in each row of the Tables (\ref{RMA}) and (\ref{RMA2}), when one condition in a special row is satisfied, other related applicable conditions in that special row must also be considered. For example, in the Table (\ref{RMA}), if a potential (or generally, any function which depends upon the scalar field) satisfies the condition $\mathfrak{B}[\varphi,\phi_{0};V(\varphi)] \geq 0$ which is in `Type-I'-row, then other `\textit{applicable}'\footnotemark[\thefndolors] conditions in that row namely $\mathfrak{B}[a,\gamma_{0};V(a)] \leq 0$, and etcetera must also be applied.\\
Indeed, the different behaviors of the scalar field, $\varphi$, specify the type of class and the different treatments of other scalar functions separate the type of sub-classes. The domains of these functions are affected by the other conditions; more precisely, there is a connection among almost all results of applying all conditions.\\
As mentioned earlier, it is more convenient to restrict ourselves to strictly monotonic scalar fields, not hybrid ones. In ref. \cite{potandro}, a set of various potentials has been listed. According to the forms of the potentials presented in this reference, it is interesting that several potentials are of monotonic cases provided that our scalar field be a strictly monotonic function, for example, the potentials of monotonic (with `time') group are:
\begin{enumerate}
	 \item The power law potentials:
\begin{align}\label{pott1}
   V(\varphi)=V_{0}{\varphi^{n}},
\end{align}
   where $n>0$ or $n<0$.
  \item The exponential potentials:
  \begin{align}\label{pott2}
  V(\varphi)&=V_{0} \exp(\lambda \varphi),\quad \text{where } \; \lambda>0 \text{ or } \lambda<0; \\ V(\varphi)&=V_{0} \left[\exp \left(\frac{\gamma}{\varphi}\right)-1 \right].
  \end{align}
  \item The unified dark matter potential:
  \begin{align}\label{pott3}
  V(\varphi)=V_{0} \left[1+\cosh^{2}(\lambda \varphi) \right].
  \end{align}
  \item The early dark energy potential:
  \begin{align}\label{pott4}
  V(\varphi)=V_{0} \left[\cosh(\lambda \varphi)-1 \right]^{p}.
  \end{align}
  \item The Chaplygin gas from the ordinary scalar field viewpoint:
  \begin{align}\label{pott5}
  V(\varphi)=V_{0} \left[\cosh(\lambda \varphi)+\frac{1}{\cosh(\lambda \varphi)} \right].
  \end{align}
\end{enumerate}
However, there are some hybrid cases, for instance:
\begin{enumerate}
	 \item The supergravity motivated potential:
\begin{align}\label{pott6}
V(\varphi)=V_{0} \frac{\exp(\varphi^2)}{\varphi^{\gamma}}.
\end{align}
  \item The pseudo-Nambu Goldstone boson potential:
\begin{align}\label{pott7}
V(\varphi)=V_{0} (1+\cos(\lambda \varphi)).
\end{align}
  \item The early dark energy potential:
\begin{align}\label{pott8}
V(\varphi)=V_{1} \exp(-\lambda_{1} \varphi) + V_{1} \exp(-\lambda_{2} \varphi),
\end{align}
this potential is hybrid when $\frac{\lambda_{1}V_{1}}{\lambda_{2}V_{2}}<0$ and $\lambda_{1}\neq \lambda_{2}$, otherwise, it is monotonic with time.
  \item The Albrecht-Skordis potential:
\begin{align}\label{pott9}
V(\varphi)=V_{0} \left[V_{1}+(\varphi-\varphi_{0})^{2} \right]\exp(-\gamma \varphi).
\end{align}
\end{enumerate}
The monotonic regions of these hybrid potentials are separated by the solution of $V^{\prime}=0$, then one may focus on monotonic parts by tuning constant parameters to stretch the evolution interval such that be monotonic in the interval of interest (Note that it is only a suggestion to make our work easy, not more). The free parameters of the aforementioned potentials can be constrained by using the current cosmological data; see refs. \cite{POT1,POT2,POT3,POT4,POT5,POT6,POT7,POT8,POT9,POT10,POT11,POT12,POT13}.\\
As a final point about the scalar-field functions, it is worth mentioning that in data analysis we can first consider the conditions, then the treatments of the potential versus time may be obtained spontaneously without imposing a special behavior. In the first example of this paper, we also act in this way.

As shown in the figure (\ref{fig7}) and table (\ref{yastable}), the treatments of the Hubble parameter is class-independent and common between both classes when it is in terms of the time, $H(t)$, scale factor, $H(a)$, and redshift, $H(z)$. In the first example of this paper, we use $H(a)$ since we have its form and hence we deal with class-independent one, but in some approaches (so-called super-potential approach) which the form of $H(\varphi)$ is accessible, the type of expansion and phase of the universe are completely class-dependent (See the last rows of the tables (\ref{RMA}) and (\ref{RMA2})). However, the inflection and phase-transition points are class-independent when one considers $H=H(\varphi)$.

As a final point, it is worthwhile mentioning that when one utilizes the approach presented in ref. \cite{ref1}, the constant scalar field is excluded, because the constant scalar field leads to a wrong physical result: `There is no expansion in the universe' (See the proof of the Theorem-1 in the supplement; `Proof \ref{SPARM1}').\\

\textcolor[rgb]{1.00,0.00,0.50}{\ding{228}}
\textcolor[rgb]{1.00,0.00,1.00}{\textbf{The prescription for applying the $\mathfrak{B}\text{-function}$ in the analysis of the exact solutions:}}

In the analysis of the exact solutions of alternative theories, we have $a(t)$, therefore, if we plot $\mathfrak{B}[t,0;H(t)]$ versus $\mathfrak{B}[t,0;a(t)]$, then by comparing it with `Plot (A)' in the figure (\ref{fig7}) and using its relevant table (i.e. Table (\ref{yastable})), the most important features of the universe under that model will easily appear. The type of class in this context is appeared spontaneously, because we know $\varphi(t)$. The rest of the procedure is similar to the previous application.\\
Since the manner of using the $\mathfrak{B}\text{-function}$ method in order to analyze the exact solutions of the extended theories of gravity is clear, hence we do not present an example to this application in this paper.\\
In data analysis and also plotting, the ranges obtained by the $\mathfrak{B}\text{-function}$ method would be so helpful.\\
\begin{table*}
\caption{The first class of the $\mathfrak{B}\text{-function}$ method: Based on the decreasing scalar field with time (D.S.F.).} \label{RMA}
\centering
\begin{tabular}{l l l c}
\toprule[1.7pt]\\[-1.9ex]
\textbf{Family of the Parameters} &\textbf{Notations} & \textbf{Conditions} &\textbf{Descriptions and Physical Meaning}
\\ [1ex]
\toprule[1.4pt]\\[-1.9ex]
 & $a=a(t)$ & $\mathfrak{B}[t,0;a]>0$ \\[-1ex]
\raisebox{1.5ex}{The scale factors} &
$\tilde{a}=a(\varphi)$ & $\mathfrak{B}[\varphi,\beta_{0};\tilde{a}]<0$& \raisebox{1.5ex}{Expansion of the universe}\\[1ex]
\toprule[0.7pt]\\[-1.9ex]
 & $\varphi=\varphi(t)$ & $\mathfrak{B}[t,0;\varphi]<0$ & \\
The scalar fields & $\grave{\varphi}=\varphi(z)$ & $\mathfrak{B}[z,1;\grave{\varphi}]>0$ & Decreasing Scalar Field w.r.t. time (D.S.F.)\\
 & $\tilde{\varphi}=\varphi(a)$ & $\mathfrak{B}[a,\alpha_{0};\tilde{\varphi}]<0$& \\[1ex]
\toprule[0.7pt]\\[-1.9ex]
 & $\mathbb{F}=F(t)$ & $\mathfrak{B}[t,0;\mathbb{F}] \leq 0$ & \\
 & $\grave{F}=F(z)$ & $\mathfrak{B}[z,1;\grave{F}] \geq 0$ & \\[-1ex]
 \raisebox{1.5ex}{Type-I of the scalar functions} &
  $F=F(\varphi)$& $\mathfrak{B}[\varphi,\phi_{0};F]\geq 0$ & \raisebox{1.5ex}{Decreasing scalar-field function w.r.t. time}\\[-1ex]
 & $\tilde{F}=F(a)$ & $\mathfrak{B}[a,\gamma_{0};\tilde{F}]\leq 0$ & \raisebox{1.5ex}{(e.g. $F=F_{0}\varphi^2$)}\\[1ex]
\toprule[0.7pt]\\[-1.9ex]
 & $\mathbb{F}=F(t)$ & $\mathfrak{B}[t,0;\mathbb{F}] \geq 0$ & \\
 & $\grave{F}=F(z)$ & $\mathfrak{B}[z,1;\grave{F}] \leq 0$ & \\[-1ex]
 \raisebox{1.5ex}{Type-II of the scalar functions} &
  $F=F(\varphi)$& $\mathfrak{B}[\varphi,\phi_{0};F]\leq 0$ & \raisebox{1.5ex}{Increasing scalar-field function w.r.t. time} \\[-1ex]
 & $\tilde{F}=F(a)$ & $\mathfrak{B}[a,\gamma_{0};\tilde{F}]\geq 0$ & \raisebox{1.5ex}{(e.g. $F=F_{0}\varphi^{-2}$)} \\[1ex]
 \toprule[0.7pt]\\[-1.9ex] %
 &  & $\mathfrak{B}[\varphi,0;\tilde{H}] <0$ & Super-accelerated expansion \\
 &  & $\mathfrak{B}[\varphi,0;\tilde{H}] < -\mathfrak{B}^{-1}[a,0,\tilde{\varphi}]$ & Accelerated expansion \\
 & & $\mathfrak{B}[\varphi,0;\tilde{H}] = -\mathfrak{B}^{-1}[a,0,\tilde{\varphi}]$ & The inflection point\\[-1ex]
 & \raisebox{1.5ex}{$\tilde{\varphi}=\varphi(a)$}& $\mathfrak{B}[\varphi,0;\tilde{H}] > -\mathfrak{B}^{-1}[a,0,\tilde{\varphi}]$ & Decelerated expansion\\[-1ex]
 \raisebox{1.5ex}{The Hubble parameter} & \raisebox{1.5ex}{$\tilde{H}=H(\varphi)$}& $\mathfrak{B}[\varphi,0;\tilde{H}]<0 $ & Phantom-like regime\\
 & & $\mathfrak{B}[\varphi,0;\tilde{H}]=0$ & Phase transition point (deSitter era/expansion)\\
 & & $\mathfrak{B}[\varphi,0;\tilde{H}]>0$ & (Quintessence/Non-Phantom)-like regime\\[1ex]

\toprule[1.7pt]\\[-1.9ex]
\end{tabular}
\end{table*}
\begin{table*}
\caption{The second class of the $\mathfrak{B}\text{-function}$ method: Based on the increasing scalar field with time (I.S.F.).} \label{RMA2}
\centering
\begin{tabular}{l l l c}
\toprule[1.7pt]\\[-1.9ex]
\textbf{Family of the Parameters} &\textbf{Notations} & \textbf{Conditions} &\textbf{Descriptions and Physical Meaning}
\\ [1ex]
\toprule[1.4pt]\\[-1.9ex]
 & $a=a(t)$ & $\mathfrak{B}[t,0;a]>0$ \\[-1ex]
\raisebox{1.5ex}{The scale factors} &
$\tilde{a}=a(\varphi)$ & $\mathfrak{B}[\varphi,\beta_{0};\tilde{a}]>0$& \raisebox{1.5ex}{Expansion of the universe}\\[1ex]
\toprule[0.7pt]\\[-1.9ex]
 & $\varphi=\varphi(t)$ & $\mathfrak{B}[t,0;\varphi]>0$ & \\
The scalar fields & $\grave{\varphi}=\varphi(z)$ & $\mathfrak{B}[z,1;\grave{\varphi}]<0$ & Increasing Scalar Field w.r.t. time (I.S.F.)\\
 & $\tilde{\varphi}=\varphi(a)$ & $\mathfrak{B}[a,\alpha_{0};\tilde{\varphi}]>0$& \\[1ex]
\toprule[0.7pt]\\[-1.9ex]
 & $\mathbb{F}=F(t)$ & $\mathfrak{B}[t,0;\mathbb{F}] \geq 0$ & \\
 & $\grave{F}=F(z)$ & $\mathfrak{B}[z,1;\grave{F}] \leq 0$ & \\[-1ex]
 \raisebox{1.5ex}{Type-I of the scalar functions} &
  $F=F(\varphi)$& $\mathfrak{B}[\varphi,\phi_{0};F]\geq 0$ & \raisebox{1.5ex}{Increasing scalar-field function w.r.t. time} \\[-1ex]
 & $\tilde{F}=F(a)$ & $\mathfrak{B}[a,\gamma_{0};\tilde{F}]\geq 0$ & \raisebox{1.5ex}{(e.g. $F=F_{0}\varphi^{2}$)}\\[1ex]
\toprule[0.7pt]\\[-1.9ex]
 & $\mathbb{F}=F(t)$ & $\mathfrak{B}[t,0;\mathbb{F}] \leq 0$ & \\
 & $\grave{F}=F(z)$ & $\mathfrak{B}[z,1;\grave{F}] \geq 0$ & \\[-1ex]
 \raisebox{1.5ex}{Type-II of the scalar functions} &
  $F=F(\varphi)$& $\mathfrak{B}[\varphi,\phi_{0};F]\leq 0$ & \raisebox{1.5ex}{Decreasing scalar-field function w.r.t. time}\\[-1ex]
 & $\tilde{F}=F(a)$ & $\mathfrak{B}[a,\gamma_{0};\tilde{F}]\leq 0$ & \raisebox{1.5ex}{(e.g. $F=F_{0}\varphi^{-2}$)}\\[1ex]
 \toprule[0.7pt]\\[-1.9ex]
 &  & $\mathfrak{B}[\varphi,0;\tilde{H}] >0$ & Super-accelerated expansion \\
 &  & $\mathfrak{B}[\varphi,0;\tilde{H}] > -\mathfrak{B}^{-1}[a,0,\tilde{\varphi}]$ & Accelerated expansion \\
 & & $\mathfrak{B}[\varphi,0;\tilde{H}] = -\mathfrak{B}^{-1}[a,0,\tilde{\varphi}]$ & The inflection point\\[-1ex]
 & \raisebox{1.5ex}{$\tilde{\varphi}=\varphi(a)$}& $\mathfrak{B}[\varphi,0;\tilde{H}] < -\mathfrak{B}^{-1}[a,0,\tilde{\varphi}]$ & Decelerated expansion\\[-1ex]
 \raisebox{1.5ex}{The Hubble parameter} & \raisebox{1.5ex}{$\tilde{H}=H(\varphi)$}& $\mathfrak{B}[\varphi,0;\tilde{H}]>0 $ & Phantom-like regime\\
 & & $\mathfrak{B}[\varphi,0;\tilde{H}]=0$ & Phase transition point (deSitter era/expansion)\\
 & & $\mathfrak{B}[\varphi,0;\tilde{H}]<0$ & (Quintessence/Non-Phantom)-like regime\\[1ex]

\toprule[1.7pt]\\[-1.9ex]
\end{tabular}
\end{table*}

\begin{table*}
\caption{The descriptions of the figure (\ref{fig7}).} \label{yastable}
\begin{scriptsize}
\centering
\begin{tabular}{p{6cm} l}
\toprule[1.7pt]\\[-1.9ex]
\textbf{Range/Line} &\textbf{General and non-general physical meanings} \\ [1ex]
\toprule[1.4pt]\\[-1.9ex]
The green area: $\Psi_{1}>0$ in `Plot (A)' of Fig. (\ref{fig7}) which is equivalent to $\mathfrak{B}[a,0;H(a)]>0$ in `Plot (B)' and also $\mathfrak{B}[z,+1;H(z)]<0$ in `Plot (C)'. &
 \begin{minipage} [t] {0.7\textwidth}
      \begin{itemize}
      \item \textcolor[rgb]{0.24,0.24,1.00}{The general physical meanings:}
        \subitem \textit{i}: Super-accelerated expansion;
        \subitem \textit{ii}: Phantom-like regime.
      \item \textcolor[rgb]{0.24,0.24,1.00}{The non-general physical meanings:}
      \subitem  This area is a phantom phase when you study the dark energy sector only. However, if the total Equation of State (EoS) parameter is less than $-1$, it is deducible that the EoS parameter of the dark energy is also less than $-1$. Currently, our universe is in this green area, therefore the phase of the universe may be taken as a phantom.
     \end{itemize}
     \end{minipage}\\ \\[1ex]
\toprule[0.7pt]\\
The blue area: Between $\Psi_{1}=0$ and $\Psi_{1}=-\Psi_{2}$ which is equivalent to $-1<\mathfrak{B}[a,0;H(a)]<0$ and also $0<\mathfrak{B}[z,+1;H(z)]<1$. &
      \begin{minipage} [t] {0.7\textwidth}
      \begin{itemize}
      \item \textcolor[rgb]{0.24,0.24,1.00}{The general physical meanings:}
            \subitem \textit{i}: Accelerated expansion;
            \subitem \textit{ii}: Non-phantom-like regime.
      \item \textcolor[rgb]{0.24,0.24,1.00}{The non-general physical meanings:}
      \subitem  This area is known as quintessence phase when you study the dark energy sector only. We may have permission to call this area as a quintessence-like regime in general.
     \end{itemize}
     \end{minipage}\\
 \\[1ex]
\toprule[0.7pt]\\
The yellow area: Between $\Psi_{2}=0$ and $\Psi_{1}=-\Psi_{2}$ which is equivalent to $\mathfrak{B}[a,0;H(a)]<-1$ and also $\mathfrak{B}[z,+1;H(z)]>1$. &
\begin{minipage} [t] {0.7\textwidth}
      \begin{itemize}
      \item \textcolor[rgb]{0.24,0.24,1.00}{The general physical meanings:}
            \subitem \textit{i}: Decelerated expansion;
            \subitem \textit{ii}: Non-phantom-like regime.
      \item \textcolor[rgb]{0.24,0.24,1.00}{The non-general physical meanings:}
      \subitem  This area is known as quintessence phase when you study the dark energy sector only. However, this terminology is true when you take the EoS parameter of quintessence phase less than $-1$. There is no consensus in the literature about the range of the quintessence phase, but the well-known range is between $0$ and $-1$ which in our figure would be equivalent to the area between $\Psi_{1}=0$ and $\Psi_{1}=(-3/2)\Psi_{2}$. Hence, this well-known candidate for dark energy indicates a transition from decelerated to accelerated expansion.
     \end{itemize}
     \end{minipage}\\
 \\[1ex]
\toprule[0.7pt]\\
The line $\Psi_{1}=0$ ($\Psi_{2}\text{-axis}$) which is equivalent to $\mathfrak{B}[a,0;H(a)]=0$ and also $\mathfrak{B}[z,+1;H(z)]=0$. &
\begin{minipage} [t] {0.7\textwidth}
      \begin{itemize}
      \item \textcolor[rgb]{0.24,0.24,1.00}{This line has several names based on physical reasons:}
            \subitem \textit{i}: The cosmological constant;
            \subitem \textit{ii}: Dark energy-dominated era;
            \subitem \textit{iii}: Vacuum fluid/energy/matter;
            \subitem \textit{iv}: de Sitter era/expansion;
            \subitem \textit{v}: Phantom divide line;
            \subitem \textit{vi}: Phase transition line.
     \end{itemize}
     \end{minipage}\\
 \\[1ex]
\toprule[0.7pt]\\
The line $\Psi_{1}=\frac{-1}{2}\Psi_{2}$ which is equivalent to $\mathfrak{B}[a,0;H(a)]=\frac{-1}{2}$ and also $\mathfrak{B}[z,+1;H(z)]=\frac{1}{2}$. &
\begin{minipage} [t] {0.7\textwidth}
      \begin{itemize}
\item The frustrated network of domain walls
\end{itemize}
     \end{minipage}\\
\\[1ex]
\toprule[0.7pt]\\
The line $\Psi_{1}=-\Psi_{2}$ which is equivalent to $\mathfrak{B}[a,0;H(a)]=-1$ and also $\mathfrak{B}[z,+1;H(z)]=1$. &
\begin{minipage} [t] {0.7\textwidth}
      \begin{itemize}
\item \textcolor[rgb]{0.24,0.24,1.00}{The general physical meanings:}
\subitem \textit{i}: The line of inflection points namely shifting from decelerated expansion to accelerated expansion;
\subitem \textit{ii}: Expansion with constant rate;
\item \textcolor[rgb]{0.24,0.24,1.00}{The non-general physical meanings:}
\subitem The frustrated network of cosmic strings.
\end{itemize}
     \end{minipage}\\
 \\[1ex]
\toprule[0.7pt]\\
The line $\Psi_{1}=\frac{-3}{2}\Psi_{2}$ which is equivalent to $\mathfrak{B}[a,0;H(a)]=\frac{-3}{2}$ and also $\mathfrak{B}[z,+1;H(z)]=\frac{3}{2}$. &
\begin{minipage} [t] {0.7\textwidth}
      \begin{itemize}
\item (Pressureless) Matter-dominated era (Known as dust-dominated era; dust is a simple and a very good candidate to the pressureless matter. However, the `Cold Dark Matter'(CDM) is also a suitable case to this end).
\end{itemize}
\end{minipage}\\
\\[1ex]
\toprule[0.7pt]\\
The line $\Psi_{1}=-2\Psi_{2}$ which is equivalent to $\mathfrak{B}[a,0;H(a)]=-2$ and also $\mathfrak{B}[z,+1;H(z)]=2$. &
\begin{minipage} [t] {0.7\textwidth}
      \begin{itemize}
\item Radiation-dominated era.
\end{itemize}
\end{minipage}\\
\\[1ex]
\toprule[0.7pt]\\
The line $\Psi_{1}=-3\Psi_{2}$ which is equivalent to $\mathfrak{B}[a,0;H(a)]=-3$ and also $\mathfrak{B}[z,+1;H(z)]=3$. &
\begin{minipage} [t] {0.7\textwidth}
      \begin{itemize}
\item Stiff-fluid-dominated era.
\end{itemize}
\end{minipage}\\
\\[1ex]
\toprule[0.7pt]\\
The area between $\Psi_{1}=\frac{-3}{2}\Psi_{2}$ and $\Psi_{1}=-3\Psi_{2}$ which is equivalent to $-3<\mathfrak{B}[a,0;H(a)]<\frac{-3}{2}$ and also $\frac{3}{2}<\mathfrak{B}[z,+1;H(z)]<3$. &
\begin{minipage} [t] {0.7\textwidth}
      \begin{itemize}
\item This area is related to the `Barotropic fluid'.
\end{itemize}
\end{minipage}\\
\\[1ex]
\toprule[0.7pt]\\
The area between $\Psi_{1}=0$ and $\Psi_{1}=\frac{-3}{2}\Psi_{2}$ which is equivalent to $\frac{-3}{2}<\mathfrak{B}[a,0;H(a)]<0$ and also $0<\mathfrak{B}[z,+1;H(z)]<\frac{3}{2}$. &
\begin{minipage} [t] {0.7\textwidth}
      \begin{itemize}
\item This area is related to the `Standard XCDM' (X-matter plus cold dark matter).
\end{itemize}
\end{minipage}\\
\\[1ex]
\toprule[1.7pt]\\[-1.9ex]
\end{tabular}
\end{scriptsize}
\end{table*}

It is important to mention that according to observational data, the present value of the common condition $\mathfrak{B}[t, 0; a(t)]>0$ is `one', viz.,
\begin{align}\label{ansherli}
\mathfrak{B}[t,0;a(t)] \bigg|_{t=t_{0}}=t_{0}\left(\frac{\dot{a}(t)}{a(t)} \right)_{t=t_{0}}
=t_{0} H_{0}=1,
\end{align}
where $t_{0}$ and $H_{0}$ represent the present time and the present amount of the mean Hubble parameter, respectively. In (\ref{ansherli}) and also from now on, a dot denotes derivative with respect to the cosmic time $t$.

For the LRS B-I (Locally  Rotationally Symmetric Bianchi type I) universe model which is given by
\begin{align}\label{line element}
\mathrm{d}s^2=\mathrm{d}t^2-A^2(t)\mathrm{d}x^2-B^2(t) \left[\mathrm{d}y^2+\mathrm{d}z^2\right],
\end{align}
where the expansion radii $A$ and $B$ are functions of time $t$, there is a well-known physical assumption $A=B^{m}$ with $m\neq 0$. Note that $m=0$ is nonphysical because it means that one of the scale factors is constant (i.e. $A=1$), and $m=1$ is FRW space-time. This relation arises from the condition that in a spatially homogeneous model the ratio of the shear scalar $\sigma$,
\begin{align}\label{zub-sig}
\sigma^2=\frac{1}{2}\sigma_{ab}\sigma^{ab}=\frac{1}{3}
\left(\frac{\dot{A}}{A}-\frac{\dot{B}}{B} \right)^2,
\end{align}
to the expansion scalar $\Theta$,
\begin{align}\label{zub-Thet}
\Theta=u^{a}_{;a}=\frac{\dot{A}}{A}+2\frac{\dot{B}}{B},
\end{align}
is constant (i.e. $\sigma / \Theta=constant$). It is worth mentioning that the common condition in both classes $\mathfrak{B}[t,0;a]=t\dot{a}/a=tH_{i}>0$ which must be held for all scale factors, in this model implies $m>0$ (because of $H_{1}=m H_{2}$) which agrees with observational limits. Therefore, unlike several papers in which negative amounts of $m$ have been allowed, we here observe that its negative amounts are completely nonphysical. Some authors interpret its negative value as an expansion on the opposite side, which is wrong. The meaning of the negative amount of $m$ is the ``contraction''. The negative amount of the `scale factor' indicates expansion on the opposite side.\\

\textcolor[rgb]{1.00,0.00,0.50}{\ding{228}} \textcolor[rgb]{1.00,0.00,1.00}{\textbf{The prescription for applying the $\mathfrak{B}\text{-function}$ in the analysis of the phase space:}}\\
I prefer to explain the application of $\mathfrak{B}\text{-function}$ in the analysis of the phase space with an example in the sub-section (\ref{yaasin}).\\ Some useful relations in order to this application are presented in what follows.\\

\noindent\pqrmybox{red}{\textbf{$\bullet$} \textbf{An important point:}}
\text{ }

However, the $\mathfrak{B}\text{-function}$ method provides two classes, but it seems that the `Class-1 (Table (\ref{RMA}))' is better than one another. In this paper, we also investigate the first example which is about the analysis of the reconstruction method, according to the `Class-1 (Table (\ref{RMA}))', namely we reconstruct the parameters based on the decreasing scalar field, because of the following statements:\\
However, there is no detectable evidence for such a scalar field. But, it is treated as a dynamical cosmological constant. As we know, cosmological constant has got a real physical meaning. It is the vacuum energy density, which is the sum of zero-point energies of quantum fields. Its value is calculated to be around $10^{74} \; \mathrm{GeV}^4$. But presently, if it exists then it is of the order of the square of the present value of the Hubble parameter ($\rho_{\Lambda} = (3/8 \pi G) H^2_0 = 10^{-47} \; \mathrm{GeV}^4$). Hence, $\rho_{\Lambda}$ is $10^{-121}$ order of magnitude less than $\rho_{\mathrm{vacuum}}$. This is possible if  we start with a large scalar field $\rho_{\varphi} = (1/2) \dot{\varphi}^2 + V(\varphi)$, and presently $\rho_{\varphi} = 10^{-47} \; \mathrm{GeV}^4$. Furthermore, for nearly 60-e fold inflation, $\varphi$ must have been decayed. Then transition from cold big bang to the hot big bang requires further decay of the scalar field through the production of particles. If at the end of inflation it monotonically decays then only the problem of the cosmological constant is evaded. And also, in many papers of alternative theories of gravity, the scalar field is of decreasing nature. Therefore, altogether it may be argued that reconstruction via decaying scalar field seems reasonable than increasing one. Hence, we prefer to proceed via the former one.\\

\noindent\pqrmybox{red}{\textbf{$\bullet$} \textbf{Other Applications of $\mathfrak{B}\text{-function}$ and some useful relations:}}
\text{ }

The family of $\mathfrak{B}$-functions may also be helpful in dynamical system approaches because of their dimensionless properties. Let us present some beneficial connectional formulas for this purpose and also for other applications (e.g. 1- inflation era; 2- when one changes the variable cosmic time $t$ to redshift $z$, or to $N=\ln(a)$). These connectional formulas may be proved so easily by the reader.
\begin{enumerate}
\item
\begin{align}\label{BCF00}
&\mathfrak{B}^{\star}[N,0;z(N)]=\frac{a_{0}}{a-a_{0}},
\end{align}
where $a_{0}$ is the present amount of the scale factor $a$.
\item
\begin{align}\label{BCF1}
\mathfrak{B}[t,0;a(t)]=-\mathfrak{B}[t,0;(1+z(t))].
\end{align}
\item
\begin{align}\label{BCF2}
\mathfrak{B}[t,0;x]=\mathfrak{B}[t,0;H(t)]+1,
\end{align}
where $x=\mathfrak{B}[t,0;a(t)]$. Indeed, eq. (\ref{BCF2}) presents the evolution of $\mathfrak{B}[t,0;a(t)]\text{-function}$.
\item A useful point in considering dynamical systems is that, generally, the evolution of $K:=\mathfrak{B}^{\star}[N,0;\mathcal{F}(N)]$, where $\mathcal{F}$ is an arbitrary function, versus $N$ is given by $\mathfrak{B}^{\star}[N,0;K]$.
\item
\begin{align}\label{BCF5}
\mathfrak{B}[t,0;\mathcal{F}(t)]=-\mathfrak{B}[t,0;a(t)] \; \mathfrak{B}[z,1;\mathcal{F}(z)],
\end{align}
where $\mathcal{F}$ is an arbitrary function.
\item
\begin{align}\label{BCF20}
\mathfrak{B}[t,0;\mathcal{F}(t)]=\mathfrak{B}[t,0;(1+z(t))] \; \mathfrak{B}[z,1;\mathcal{F}(z)].
\end{align}
\item
\begin{align}\label{BCF3}
\frac{\mathfrak{B}[t,0;\mathcal{F}(t)]}{\mathfrak{B}[t,0;a(t)]}
=\frac{\mathfrak{B}[N,0;\mathcal{F}(N)]}{N}=\mathfrak{B}^{\star}[N,0;\mathcal{F}(N)].
\end{align}
\item
\begin{align}\label{BCF6}
\mathfrak{B}^{\star}[N,0;\mathcal{F}(N)]= -\mathfrak{B}[z,1;\mathcal{F}(z)].
\end{align}
\item
\begin{align}\label{BCF7}
\mathfrak{B}[a,0;H(a)]= - \mathfrak{B}[z,1;H(z)].
\end{align}
\item
\begin{align}\label{BCF0007}
\mathfrak{B}[a,0;H(a)]= \mathfrak{B}^{\star}[N,0;H(N)].
\end{align}
\item The closed form of the well-known relation $z+1=(a_{0}/a)$:
\begin{align}\label{BCF8}
\mathfrak{B}[z,1;a(z)]= -1.
\end{align}
\item
\begin{align}\label{BCF88}
\mathfrak{B}[(1+z),0;a(1+z)]= -1.
\end{align}
\item If $P=\rho \omega$, then
\begin{align}\label{BCF4}
\omega &=\frac{-1}{3}\frac{\mathfrak{B}[t,0;\rho(t)]}{\mathfrak{B}[t,0;a(t)]}-1
=\frac{-1}{3} \frac{\mathfrak{B}[N,0;\rho(N)]}{N}-1 \nonumber \\&=\frac{1}{3}\mathfrak{B}[z,1;\rho(z)]-1.
\end{align}
\item There is a general chain formula which will be useful when changing the variables is demanded:
\begin{equation}\label{Chain rule}
\frac{\mathfrak{B}[u,c;\mathcal{F}(u)]}{\mathfrak{B}[u,c;y(u)]}
=\mathfrak{B}[y,0;\mathcal{F}(y)]
\end{equation}
where $u$, $y$, and $\mathcal{F}$ are arbitrary and especially nonzero variable/function, and  $c$ is a constant.
\item The constant member of the argument of the $\mathfrak{B}\text{-function}$ may be moved to zero by the general relation:
\begin{equation}\label{p389}
\mathfrak{B}[u,c,\mathcal{F}(u)]=\mathfrak{B}[(u+c),0;\mathcal{F}(u+c)].
\end{equation}
\item It is not hard to show for FRW case that, in general, there is a relation between $\mathfrak{B}[t,0;a(t)]$ and the curvature $R$ as
\begin{equation}\begin{split}\label{BCF10}
\mathfrak{B}^{n+2}[t,0;a(t)]=f_{n+2}(q,j,s,\cdots)\; t^{n+2}\; \frac{\mathrm{d}^{n} R}{\mathrm{d}t^{n}}; \quad n\geq 0,
\end{split}\end{equation}
where $f_{n+2}(q,j,s,\cdots)$ is a function of the mean deceleration, $q$, jerk, $j$, snap, $s$, and other dimensionless parameters which appear when one writes the Taylor expansion of the mean scale factor down,
\begin{align}\label{Taylora}
a(t)=a_{0}\bigg\{&1+H_{0}(t-t_{0})
-\frac{1}{2} q_{0} H^2_{0}(t-t_{0})^2
\nonumber\\&+\frac{1}{6} j_{0} H^3_{0}(t-t_{0})^3
+\frac{1}{24} s_{0} H^4_{0}(t-t_{0})^4+\cdots \bigg\},
\end{align}
and they are defined as \cite{jjj1,jjj2,jjj3,jjj4}
\begin{equation}\begin{split}\label{Taylora1}
q&=\frac{-\ddot{a}}{a}\left(\frac{\dot{a}}{a} \right)^{-2}, \quad
j=\frac{\dddot{a}}{a}\left(\frac{\dot{a}}{a} \right)^{-3}, \\
s&=\frac{\ddddot{a}}{a}\left(\frac{\dot{a}}{a} \right)^{-4}, \qquad \cdots.
\end{split}\end{equation}
For example, for $n=0,1,$ and $2$ one has:
\begin{equation}\begin{split}\label{CURVATURE1}
\mathfrak{B}^{2}[t,0;a(t)]=f_{2}\; t^{2} \; R; \quad f_{2}=\left[6\epsilon(1-q)\right]^{-1},
\end{split}\end{equation}
\begin{equation}\begin{split}\label{CURVATURE2}
\mathfrak{B}^{3}[t,0;a(t)]=f_{3}\; t^{3} \; \dot{R}; \quad f_{3}=\left[6\epsilon(j-q-2)\right]^{-1},
\end{split}\end{equation}
\begin{equation}\begin{split}\label{CURVATURE3}
\mathfrak{B}^{4}[t,0;a(t)]=f_{4}\; t^{4} \; \ddot{R}; \quad f_{3}=\left[6\epsilon\left(s+q^{2}+8q+6 \right)\right]^{-1},
\end{split}\end{equation}
respectively. In eqs. (\ref{CURVATURE1})-(\ref{CURVATURE3}), $\epsilon$ is $+1$ and $-1$ for FRW metric signatures $(-+++)$ and $(+---)$, respectively.
\item Consequently, there is a relation among the evolution of $H$, and also its derivatives ($\dot{H},\ddot{H},\cdots$) and $\mathfrak{B}[t,0;a(t)]$:
\begin{equation}\begin{split}\label{CURVATURE4}
\mathfrak{B}\left[t,0,\frac{\mathrm{d}^{p}H(t)}{\mathrm{d}t^{p}}\right]
=\tilde{f}_{p}(q,j,s,\cdots) \mathfrak{B}[t,0;a(t)]; \quad p\geq 0,
\end{split}\end{equation}
where $\tilde{f}_{p}(q,j,s,\cdots)$ is a dimensionless function like $f_{n+2}(q,j,s,\cdots)$ but they are not the same. For example, for $p=0,1$, and $2$ we have:
\begin{equation}\begin{split}\label{CURVATURE5}
&\mathfrak{B}[t,0;H(t)]=\tilde{f}_{0}\; \mathfrak{B}[t,0;a(t)]; \\& \tilde{f}_{0}=-(q+1),
\end{split}\end{equation}
\begin{equation}\begin{split}\label{CURVATURE6}
&\mathfrak{B}[t,0;\dot{H}(t)]=\tilde{f}_{1}\; \mathfrak{B}[t,0;a(t)]; \\& \tilde{f}_{1}=\frac{3q+j+2}{-(q+1)},
\end{split}\end{equation}
\begin{equation}\begin{split}\label{CURVATURE7}
&\mathfrak{B}[t,0;\ddot{H}(t)]=\tilde{f}_{2}\; \mathfrak{B}[t,0;a(t)]; \\& \tilde{f}_{2}=\frac{s-4j-3 q^{2}-12q-6}{j+3q+2},
\end{split}\end{equation}
\item According to (\ref{BCF3}) and (\ref{BCF6}), eq. (\ref{CURVATURE4}) may be rewritten as
\begin{equation}\begin{split}\label{CURVATURE8}
&\mathfrak{B}^{\star}\left[N,0,\frac{\mathrm{d}^{p}H(N)}{\mathrm{d}N^{p}}\right]
=\frac{\mathfrak{B}\left[N,0,\frac{\mathrm{d}^{p} H(N)}{\mathrm{d}N^{p}}\right]}{N}
\\&=-\mathfrak{B}\left[z,1,\frac{\mathrm{d}^{p}H(z)}{\mathrm{d}z^{p}}\right]
=\tilde{f}_{p}(q,j,s,\cdots); \quad p \geq 0.
\end{split}\end{equation}
Note that $N=\ln(a)$ is a dimensionless parameter.
\item A well-defined function, viz.,
\begin{align}\label{take1}
\Gamma=\frac{VV^{\prime \prime}}{\left(V^{\prime}\right)^2}
\end{align}
where $V=V(\varphi)$ is a potential and a prime denotes derivative with respect to the scalar field $\varphi$, has been suggested \cite{tracker} to understand the cosmological dynamics of the potentials. The properties of $\Gamma$ determine whether the tracking solutions exist or not. For any form of the potential, three different solutions are possible:
\begin{align}\label{take2}
\left\{
  \begin{array}{ll}
    \Gamma<1, & \hbox{Thawing;} \nonumber \\
    \Gamma=1, & \hbox{Scaling;} \nonumber \\
    \Gamma>1, & \hbox{Tracker.}
  \end{array}
\right.
\end{align}
It is worth mentioning that the function (\ref{take1}) may also be expressed in terms of $\mathfrak{B}\text{-function}$ as
\begin{equation}\label{tajahmadb}
\Gamma=\frac{\mathfrak{B}[\varphi,0;V^{\prime}(\varphi)]}
{\mathfrak{B}[\varphi,0;V(\varphi)]}.
\end{equation}
\item Pursuant to the condition $A=B^{m}$ for the LRS B-I background geometry (\ref{line element}), one may easily show the following connectional formulas:
\begin{align}
\mathfrak{B}[A,0;H_{1}(A)] &=m \; \mathfrak{B}[B,0;H_{2}(B)],                      \label{w01w} \\
\mathfrak{B}[a,0;H(a)] &=\left(\frac{3}{m+2}\right) \; \mathfrak{B}[B,0;H_{2}(B)]. \label{w02w}
\end{align}
where $a$ and $H(a)$ are the mean scale factor and Hubble parameter, respectively.
\item It is interesting to mention that the $\mathfrak{B}\text{-function}$ may also appear in other eras as well, for example, the four slow-roll parameters  \cite{inflation01,inflation02}
\begin{align}
\epsilon_{1}&=\frac{\dot{H}}{H^2}=-\epsilon_{H},\\
\epsilon_{2}&=\frac{\ddot{\varphi}}{H\dot{\varphi}}=-\eta_{H},\\
\epsilon_{3}&=\frac{\dot{F}}{2HF}; \quad F=F(\varphi,R)=\frac{\partial f}{\partial R},\\
\epsilon_{4}&=\frac{\dot{E}}{2HE}; \quad E=F\left[\omega+\frac{2\dot{F}^2}{2 \dot{\varphi}^2 F} \right],
\end{align}
which are necessary to be introduced in the slow-roll regime of scalar-tensor inflation arisen from the action
\begin{align}
S=\int \left[\frac{1}{2}f(\varphi,R)-\frac{\omega(\varphi)}{2} \nabla^{c}\varphi \nabla_{c} \varphi-V(\varphi) \right] \sqrt{-g}\; \mathrm{d}^4 x,
\end{align}
may also be expressed as
\begin{align}
\epsilon_{1}&= \; \; \mathfrak{B}^{\star}[N,0;H],\\
\epsilon_{2}&= \; \; \mathfrak{B}^{\star}[N,0;\dot{\varphi}],\\
\epsilon_{3}&= \frac{1}{2}\mathfrak{B}^{\star}[N,0;F],\\
\epsilon_{4}&= \frac{1}{2} \mathfrak{B}^{\star}[N,0;\alpha],
\end{align}
respectively. Therefore, utilizing observational data, this approach may also be developed to formulate the analysis of the early inflation era as well. Though, we do not carry it out here, because it may easily be performed.
\item And, finally three trivial relations
\begin{align}
\mathfrak{B}[x,c;(\mathcal{F}_{1} / \mathcal{F}_{2})]
&=\mathfrak{B}[x,c;\mathcal{F}_{1}]
-\mathfrak{B}[x,c;\mathcal{F}_{2}],\\
\mathfrak{B}[x,c;\mathcal{F}_{1}\mathcal{F}_{2}]
&=\mathfrak{B}[x,c;\mathcal{F}_{1}]
+\mathfrak{B}[x,c;\mathcal{F}_{2}],\\
\mathfrak{B}[x,c;\mathcal{F}^{p}]
&=p \; \mathfrak{B}[x,c;\mathcal{F}].
\end{align}
where $\mathcal{F}_{i}$ are arbitrary functions in terms of $x$, and $p$ is a constant.
\end{enumerate}

\noindent\hrmybox{}{\section{Examples}}\vspace{5mm}

In order to clarify the $\mathfrak{B}\text{-function}$ method, two examples are considered in this section; one for its application in the reconstruction method, and one for its application in dynamical systems.\\

\noindent\rmybox{}{\subsection{Example 1: Application in the analysis of the reconstruction method.\label{example114}}}\vspace{5mm}

This example is related to the application of the $\mathfrak{B}\text{-function}$ method in the analysis of the reconstruction method. We pick up it from ref. \cite{ref1}. In this example, we use the notations of ref. \cite{ref1} with the differences that, in order to match with the current paper, we change two of their notations to our notations: We show the scalar field and Hubble parameter by $\varphi$ and $H$ instead of $\sigma$ and $h$, respectively. For better understanding this example, we may refer the readers to study ref. \cite{ref1}.\\
As a first example, let us consider the `example 3.1' in ref. \cite{ref1} in which the `Barotropic fluid' has been studied via the reconstruction method.\\
They studied a well-known action of the form
\begin{align}\label{AYK3}
S=\int \left[U(\varphi)R-\frac{1}{2}\varphi_{,\mu}\varphi_{,\nu}+V(\varphi)\right]\sqrt{-g}\; \mathrm{d}^4 x
\end{align}
where a scalar field non-minimally coupled to gravity, in a Friedmann-Robertson-Walker (FRW) flat spacetime. By the use of the reconstruction method, they obtained the following results for cosmological evolutions driven by a barotropic fluid with the equation of state $P=w\rho$ (where $0<w<1$):
\begin{align}
&U(\varphi)=\frac{1}{2}\gamma \varphi^2, \quad
H(a)=\frac{h_{0}}{a^{3(w+1)/2}}, \label{Example1-1} \\ &\varphi(a)=a^{\frac{p_{1,2}}{\beta}}, \quad a(\varphi)=\varphi^{\frac{\beta}{p_{1,2}}}, \label{Example1-11} \\
&V_{1,2}(\varphi)=V_{0} \left[\frac{6\gamma \beta^{2}+\alpha \beta+(\alpha-1+12 \gamma \beta)p_{1,2}}{2\beta^{2}} \right] \varphi^{\frac{2(\alpha \beta+p_{1,2})}{p_{1,2}}}, \label{Example1-111}
\end{align}
where $\gamma$, $h_{0}$, and $V_{0}$ are constants and
\begin{align}
\alpha&=\frac{-3}{2}(w+1), \quad \beta=\frac{1+4\gamma}{2\gamma}, \label{es1} \\
p_{1,2}&=\frac{1-\alpha}{2} \pm \sqrt{\left(\frac{1-\alpha}{2} \right)^2-\alpha \beta}. \label{es11}
\end{align}
Consequently, one may easily reach
\begin{align}
&H(z)=h_{0} (1+z)^{\frac{3(1+w)}{2}}, \quad
H(\varphi)=\frac{h_{0}}{\varphi^{\frac{3\beta(1+w)}{2 p_{1,2}}}},  \label{es2} \\
&V_{1,2}(a)=V_{01,02}\; a^{\frac{2(\alpha \beta+p_{1,2})}{\beta}},
\label{es21} \\
&V_{1,2}(z)=V_{01,02}\; (1+z)^{\frac{-2(\alpha \beta+p_{1,2})}{\beta}},\label{es22} \\
&U(a)=\frac{1}{2}\gamma \; a^{\frac{2 p_{1,2}}{\beta}}, \quad U(z)=\frac{1}{2}\gamma \; (1+z)^{\frac{-2 p_{1,2}}{\beta}},\label{es23} \\
&\varphi(z)=(1+z)^{\frac{-p_{1,2}}{\beta}}, \label{es24}
\end{align}
where
\begin{align}\label{farfar}
V_{01,02}=V_{0} \left[\frac{6\gamma \beta^{2}+\alpha \beta+(\alpha-1+12 \gamma \beta)p_{1,2}}{2\beta^{2}} \right].
\end{align}
As mentioned earlier, the analysis of this paper would be depended on the `Class-1'. Hence, according to Table (\ref{RMA}), we carry out it as follows:\\
\textcolor[rgb]{0.00,0.50,0.00}{\textleaf} Conditions for the Hubble parameter:\\
Using (\ref{Example1-1}), one arrives at
\begin{align}\label{behzad1}
&\mathfrak{B}[a,0;H(a)]=\frac{-3}{2}(1+w)  \nonumber \\ &\xrightarrow{\; \; 0<w<1 \; \;} -3<\mathfrak{B}[a,0;H(a)]<\frac{-3}{2}.
\end{align}
Therefore, according to the `Plot (B)' in figure (\ref{fig7}) and table (\ref{yastable}), (\ref{behzad1}) demonstrates `decelerated expansion' and `quintessence phase'. Note that the special case $w=1/3$ is the radiation dominated era in which the universe had both aforementioned properties.\\
However, we do not have an access to the form of the Hubble parameter in terms of the time through this type of reconstruction method, nonetheless, talking about its qualitative behavior is feasible. Pursuant to the figure (\ref{fig7}), table (\ref{yastable}), and both aforementioned properties, it may easily be deduced that for this model at any time one has:
\begin{align}
&\mathfrak{B}[t,0;H(t)]<0,  \label{behzad2-1}\\
&\mathfrak{B}[t,0;a(t)]< - \mathfrak{B}[t,0;H(t)],  \label{behzad2-2}
\end{align}
yielding
\begin{align}
&\dot{H}(t)<0, \label{behzad2-3}\\
&\frac{\mathrm{d}\ln(H^{-1}(t))}{\mathrm{d}t}< H(t) \quad \text{or equivalently} \quad H^2(t) < - \dot{H}(t). \label{behzad2-4}
\end{align}
Hence, the amount of Hubble parameter in this model remains greater than the rate of the Naperian logarithm of the inverse Hubble parameter---this indicates that $H(t)$ monotonically decays. Indeed, both (\ref{behzad2-3}) and (\ref{behzad2-4}) clearly demonstrate that the speed of the expansion of the universe filled with barotropic fluid goes down (i.e. decelerated expansion). Note that the condition (\ref{behzad2-3}) is due to the quintessence phase, nonetheless, it also shows decelerated expansion, hence, it hiddenly states that the decelerated expansion exists in the quintessence phase.

Pursuant to two founded properties, namely `decelerated expansion' and `quintessence phase', and Table (\ref{RMA}), the conditions
\begin{align}
&\mathfrak{B}[\varphi,0;H(\varphi)]>\frac{-1}{\mathfrak{B}[a,0;\varphi(a)]}, \label{behzad3-1} \\
&\mathfrak{B}[\varphi,0;H(\varphi)]>0,\label{behzad3-2}
\end{align}
must be held. They lead to
\begin{align}
&\frac{-3\beta(1+w)}{2p_{1,2}}> \frac{-\beta}{p_{1,2}}, \label{behzad4-1} \\
&\frac{-3\beta(1+w)}{2p_{1,2}}>0,\label{behzad4-2}
\end{align}
respectively. Since $0<w<1$, hence (\ref{behzad4-2}) gives
\begin{align}\label{behzad5}
\frac{\beta}{p_{1,2}}<0.
\end{align}
Therefore, the first condition, (\ref{behzad4-1}), is satisfied automatically:
\begin{align}\label{behzad6}
\underbrace{\frac{3}{2}(1+w)}_{\frac{3}{2}<\text{between}<3}
\underbrace{\left(\frac{-\beta}{p_{1,2}}\right)}_{>0}
<\underbrace{\left(\frac{-\beta}{p_{1,2}}\right)}_{>0}.
\end{align}

Utilizing (\ref{es2}), $\mathfrak{B}[z,1;H(z)]$ would be
\begin{align}\label{behzad6-1}
&\mathfrak{B}[z,1;H(z)]=\frac{3}{2}(1+w) \nonumber \\
&\xrightarrow{\; \; 0<w<1 \; \;} \frac{3}{2}<\mathfrak{B}[z,1;H(z)]<3,
\end{align}
which according to the `Plot (C)' in figure (\ref{fig7}) and table (\ref{yastable}), indicates two aforementioned properties, namely `decelerated expansion' and `quintessence phase'. As we observe, all applicable conditions for the Hubble parameter confirm each other.\\
\textcolor[rgb]{0.00,0.50,0.00}{\textleaf} Conditions for the scale factor and scalar field:\\
Utilizing (\ref{Example1-11}) and (\ref{es24}) we obtain
\begin{align}
&\mathfrak{B}[\varphi,0;a(\varphi)]<0 \longrightarrow \frac{\beta}{p_{1,2}}<0.\label{behzad7}\\
&\mathfrak{B}[a,0;\varphi(a)]<0 \longrightarrow \frac{\beta}{p_{1,2}}<0.\label{behzad8}\\
&\mathfrak{B}[z,1;\varphi(z)]>0 \longrightarrow \frac{-p_{1,2}}{\beta}>0.\label{behzad9}
\end{align}
which agree with (\ref{behzad5}).\\
\textcolor[rgb]{0.00,0.50,0.00}{\textleaf} Conditions for the coupling function:\\
Using (\ref{Example1-1}), $\mathfrak{B}[\varphi,0;U(\varphi)]$ turns out to be
\begin{align}\label{behzad10}
\mathfrak{B}[\varphi,0;U(\varphi)]=2.
\end{align}
It means that $U(\varphi)$ is a decreasing coupling function with time at all times (i.e. $\mathfrak{B}[t,0;U(t)]\leq 0$). Since we restrict ourselves to strictly decreasing monotonic scalar field, hence, this property is due to the fact that the $U(\varphi)$ has a monotonic form, $U(\varphi)=(1/2)\;\gamma \; \varphi^2$.\\
Because (\ref{behzad10}) satisfies the condition $\mathfrak{B}[\varphi,0;U(\varphi)] \geq 0$, therefore we must have:
\begin{align}
&\mathfrak{B}[a,0;U(a)] \leq 0 \longrightarrow \frac{2 p_{1,2}}{\beta} \leq 0, \label{behzad11-1} \\
&\mathfrak{B}[z,1;U(z)] \geq 0  \longrightarrow \frac{-2 p_{1,2}}{\beta} \geq 0. \label{behzad11-2}
\end{align}
which are compatible with the previous condition (\ref{behzad9}).\\
Before proceeding, let us extract the common domains of all conditions which studied above. Limpidly, they lead to
\begin{align}
\beta \neq 0, \quad p_{1}\neq 0, \quad p_{2} \neq 0, \label{behzad12-1}\\
\frac{\beta}{p_{1}}<0, \quad \frac{\beta}{p_{2}}<0, \label{behzad12-2}
\end{align}
together with another condition which arises from (\ref{es1}) and the assumption $0<w<1$:
\begin{align}
-3<\alpha<\frac{-3}{2}. \label{behzad12-3}
\end{align}
In the next stage, these conditions are combined to extract final results.\\
\textcolor[rgb]{0.00,0.50,0.00}{\textleaf} Conditions for the potential:\\
Fortunately, the potential in terms of the scalar field is a monotonic function (see (\ref{Example1-111}) and (\ref{pott1})); it makes our work easy since in throughout time interval of interest, it would be monotonically decreasing or increasing with time. By the use of (\ref{Example1-111}), one has:
\begin{align}\label{behzad13}
\mathfrak{B}[\varphi,0;V(\varphi)]=2 \alpha \left(\frac{\beta}{p_{1,2}} \right)+1.
\end{align}
which satisfies the condition $\mathfrak{B}[\varphi,0;V(\varphi)] \geq 0$ because of (\ref{behzad12-3}) and (\ref{behzad12-2}). Therefore, the potential strictly decays as the universe ages (i.e. $\mathfrak{B}[t,0;V(t)] \leq 0$) and can never be increasing versus time. This is due to the monotonic form of the potential in terms of the scalar field and our main assumption (i.e. Class-1; strictly decreasing scalar field). The other related conditions like $\mathfrak{B}[a,0;V(a)] \leq 0$ cannot add further thing because it is held automatically:
\begin{align}\label{behzad14}
\mathfrak{B}[a,0;V(a)]= \underbrace{2\alpha}_{\substack{\text{$<0$;}\\ \text{cf. (\ref{behzad12-3})}}}+\underbrace{\frac{p_{1,2}}{\beta}}_{\substack{\text{$<0$;}\\ \text{cf. (\ref{behzad12-1})-(\ref{behzad12-2})}}}.
\end{align}

\textcolor[rgb]{0.82,0.00,0.00}{\textleaf} \textbf{Common domains:}\\
According to (\ref{behzad12-2}), there are two possibilities:
\begin{align}\label{behzad15}
\left\{
  \begin{array}{ll}
    \beta<0,\; \text{ and }\; \{p_{1}>0, p_{2}>0\}; \\
    \text{Or},\\
    \beta>0,\; \text{ and }\; \{p_{1}<0, p_{2}<0\}.
  \end{array}
\right.
\end{align}
Now, let us consider both of these possibilities:

\textcolor[rgb]{1.00,0.00,0.50}{\ding{52}} The former one condition, $\beta<0,\; \text{ and }\; \{p_{1}>0, p_{2}>0\}$:\\
Using (\ref{es1}), $\beta<0$ leads to a condition for $\gamma$:
\begin{align}\label{behzad16}
 \frac{-1}{4} < \gamma <0.
\end{align}
Furthermore, using (\ref{es11}) and (\ref{behzad12-3}), it is revealed that both $p_{1}$ and $p_{2}$ can be a positive real number as we want, provided that we impose
\begin{align}\label{behzad17}
\frac{-25}{48}<\beta<0,
\end{align}
due to the squared term in $p_{1,2}$. By the use of (\ref{es1}) and (\ref{behzad17}), one may obtain a range for $\gamma$:
\begin{align}\label{behzad17-1}
-1<\gamma<\frac{-24}{121}.
\end{align}
As already noticed, the common domains of all conditions must be adopted, hence, by the combination of both (\ref{behzad16}) and (\ref{behzad17-1}) we arrive at the final range for $\gamma$:
\begin{align}\label{behzad17-2}
\frac{-1}{4}<\gamma<\frac{-24}{121}.
\end{align}
which leads to the previous range obtained in (\ref{behzad17}) for $\beta$. We anticipate this range for $\beta$ because for $\beta$ there were two conditions $\beta<0$ and (\ref{behzad17}), so the common domain would be again (\ref{behzad17}).\\

\textcolor[rgb]{1.00,0.00,0.50}{\ding{52}} The latter one condition, $\beta>0,\; \text{ and }\; \{p_{1}<0, p_{2}<0\}$:\\
In this case, $\beta>0$ would be satisfied in two ways
\begin{align}\label{behzad18}
\left\{
  \begin{array}{ll}
    \gamma >0; \\
    \text{or},\\
    \gamma < \frac{-1}{4},
  \end{array}
\right.
\end{align}
where we have used (\ref{es1}).\\
Unlike $p_{1}$ that, in this case, is positive, $p_{2}$, viz,
\begin{align}
p_{2}=\frac{1-\alpha}{2} - \sqrt{\left(\frac{1-\alpha}{2} \right)^2-\alpha \beta},
\end{align}
remains negative which is our objective, hence only $p_{2}$ is acceptable and consequently, only $V_{2}(\varphi)$ would be true. In this case, there is no bound on $\beta$ since the squared term is positive because of (\ref{behzad12-3}) and $\beta>0$.\\

\textcolor[rgb]{1.00,0.00,0.50}{\ding{224}} Therefore, in a nutshell, the final conditions are extracted as follows:\\
\begin{enumerate}
\item The first set:
\begin{align}\label{behzad19}
\left\{
\begin{array}{l}
-3<\alpha<\frac{-3}{2}; \\ \\
\frac{-1}{4}<\gamma<\frac{-24}{121} \text{ or equivalently } \frac{-25}{48}<\beta<0;\\ \\
\text{Both $p_{1}$ and $p_{2}$ are acceptable and $p_{1,2}>0$}.
\end{array}
\right.
\end{align}
\item The second set:
\begin{align}\label{behzad19-1}
\left\{
  \begin{array}{l}
    -3<\alpha<\frac{-3}{2}; \\ \\
     \gamma>0 \text{ or equivalently } \beta>0;\\ \\
    \text{Only $p_{2}$ is acceptable and $p_{2}<0$}.
  \end{array}
\right.
\end{align}
\item The third set:
\begin{align}\label{behzad19-2}
\left\{
  \begin{array}{l}
    -3<\alpha<\frac{-3}{2}; \\ \\
    \gamma< \frac{-1}{4} \text{ or equivalently } \beta>0;\\ \\
    \text{Only $p_{2}$ is acceptable and $p_{2}<0$}.
  \end{array}
\right.
\end{align}
\end{enumerate}
One of the above conditions is sufficient to give us the desired results. Each of the three sets yields a scalar field, a potential, and a coupling function which strictly decay versus time. The important behaviors of the universe in this case, namely `decelerated expansion' and `quintessence phase', are common among them.\\
Furthermore, pursuant to the three sets, and (\ref{take1}) which reads
\begin{align}\label{behzad20}
\Gamma=1-\frac{p_{1,2}}{2(\alpha \beta + p_{1,2})}
\end{align}
for the potentials (\ref{Example1-111}), only `Thawing' solutions, $\Gamma<1$, are possible and since (\ref{behzad20}) is time-independent, hence, both $V_{1,2}(\varphi)$ maintain this property in all times of interest.\\

\noindent\rmybox{}{\subsection{Example 2: Application in dynamical systems.\label{yaasin}}}\vspace{5mm}

In this example, indicating the application of our new method in dynamical systems (especially phase space) is our objective. Indeed, we want to show that in what manner one can convert the usual system to a system in terms of $\mathfrak{B}\text{-function}$. For this reason, we ignore to give a complete study here. The remains may be done easily by the reader. \\

Suppose that studying the phase space of flat FRW model derived from scalar-tensor gravity, viz,
\begin{align}\label{example21}
S=\int \left[f(\varphi)R+\frac{1}{2}\varphi_{,\mu}\varphi_{,\nu}-V(\varphi)\right]\sqrt{-g}\; \mathrm{d}^4 x,
\end{align}
is our objective. A very good study of this action has been presented in ref. \cite{carsallv}. As we know, in order to analyze the phase space of this context, the field equations need to be recast in dynamical system form. In FRW metric with the signature $(+,-,-,-)$, the field equations reduce to
\begin{align}
\frac{\ddot{a}}{a}+H\frac{\dot{\varphi}}{\varphi}+\frac{\ddot{\varphi}}{\varphi}
+\left(1-\frac{1}{6f_{0}} \right)\frac{\dot{\varphi}^2}{\varphi^2}+\frac{V_{0}}{6f_{0}}\varphi^{n-2}=0, \label{example22-1}\\
\frac{\ddot{\varphi}}{\varphi}+3H \frac{\dot{\varphi}}{\varphi}+12f_{0}\frac{\ddot{a}}{a}+12f_{0}\frac{\dot{a}^2}{a^2}
+nV_{0} \varphi^{n-2}=0, \label{example22-2}\\
H^2 +2H\frac{\dot{\varphi}}{\varphi}+\frac{V_{0}}{6f_{0}}\varphi^{n-2} +\frac{1}{12f_{0}}\frac{\dot{\varphi}^2}{\varphi}=0, \label{example22-3}
\end{align}
where the first is the Euler-Lagrange equation for the scale factor $a$, the second is the Hamiltonian constraint, i.e. the $\binom{0}{0}$-Einstein equation, and the third is the Euler-Lagrange equation for the scalar field $\varphi$, namely the Klein-Gordon equation. We have assumed that the nonminimal coupling is $f(\varphi)=f_{0}\varphi^2$ and the effective potential is $V(\varphi)=V_{0}\varphi^{n}$, in the equations above.\\
As usual, the equations (\ref{example22-1})-(\ref{example22-3}) may be converted into an autonomous system of first-order differential equations by defining the following set of expansion normalized variables:
\begin{align}\label{example23}
x=\frac{\dot{\varphi}}{\varphi H}, \quad y=\frac{V_{0}\varphi^{n-2}}{6f_{0}H^2}.
\end{align}
Now, $x$ may be written as
\begin{align}\label{example25}
x=\frac{\mathfrak{B}[t,0;\varphi(t)]}{\mathfrak{B}[t,0;a(t)]}
=\mathfrak{B}^{\star}[N,0;\varphi(N)].
\end{align}
As is observed, $x$ is dimensionless, hence, $y$ would also be too, and may be written down as:
\begin{align}\label{example26}
y=\frac{V_{0} \varphi^{n-2}\; t^{2}}{6f_{0}} \frac{1}{\mathfrak{B}^{2}[t,0;a(t)]}.
\end{align}
Obviously, the term $V_{0} \varphi^{n-2}\; t^{2}/ (6f_{0})$ in (\ref{example26}) is dimensionless. If you define a new function $G$ as
\begin{align}\label{example27}
G= c_{1} \exp \left[\int  \; \sqrt{\frac{V_{0}}{6f_{0}} \varphi^{n-2}}\; \mathrm{d}t \right],
\end{align}
or
\begin{align}\label{example27-1}
G= c_{1} \exp \left[- \int \; \sqrt{\frac{V_{0}}{6f_{0}} \varphi^{n-2}}\; \mathrm{d}t \right],
\end{align}
where $c_{1}$ is a constant, then $y$ turns out to be
\begin{align}\label{example28}
y={\mathfrak{B}^{\star}}^{2}[N,0;G(N)],
\end{align}
and therefore, our system will be regarded as an autonomous system and the equations will take a simple form.\\
Obviously, the evolution of the system and the critical points will then be given at the coordinate $(x,y)$ where
\begin{align}\label{example29}
x=\mathfrak{B}^{\star}[N,0;\varphi(N)] , \quad y={\mathfrak{B}^{\star}}^{2}[N,0;G(N)].
\end{align}
Generally, the evolution of $K:=\mathfrak{B}^{\star}[N,0;\mathcal{F}(N)]$, where $\mathcal{F}$ is an arbitrary function, versus $N$ is given by $\mathfrak{B}^{\star}[N,0;K]$. Therefore, using the variables in (\ref{example29}), we obtain
\begin{align}\label{example30-1}
\mathfrak{B}^{\star}[N,0;x]=&\frac{1}{(12f_{0}-1)x}\biggl[12f_{0}+2(1+6f_{0})x
 \nonumber \\&+(5-24f_{0})x^2-\left(1-\frac{1}{6f_{0}} \right)x^3\nonumber \\&+6f_{0}(n-2)y-(1-6nf_{0})xy \biggl],
\end{align}
\begin{align}\label{example30-2}
\mathfrak{B}^{\star}[N,0;y]=&\frac{2}{12f_{0}-1}\biggl[24f_{0}
-1+4x \nonumber \\ &+(12f_{0}-1)(n-2)x- \left(1-\frac{1}{6f_{0}} \right)x^2 \nonumber \\ &-(1-6nf_{0})y \biggl].
\end{align}
The dynamical variables are constrained by
\begin{align}\label{example31}
1+2x+\frac{1}{12f_{0}}x^{2}+y=0.
\end{align}
The associated phase space is two-dimensional and the evolution of the system is constrained by (\ref{example31}).\\
Indeed, in the usual way we use the coordinate $(x,y)$, where $x=\dot{\varphi}/(\varphi H)$ and $y=V_{0}\varphi^{n-2}/(6f_{0}H^2)$, and their evolutions are given by $\mathrm{d}x/\mathrm{d}N$ and $\mathrm{d}y/\mathrm{d}N$, while in our new method we use the coordinate $(x,y)$, where $x=\mathfrak{B}^{\star}[N,0;\varphi(N)]$ and $y={\mathfrak{B}^{\star}}^{2}[N,0;G(N)]$, and the evolution of the system is generated by $\mathfrak{B}^{\star}[N,0;x]$ and $\mathfrak{B}^{\star}[N,0;y]$.
The benefit of our taken procedure is that when one utilizes $\mathfrak{B}\text{-function}$ then the analysis of phase space will be convenient. For example, in the phase plane related to this example, the scalar field is decreasing scalar function when $x=\mathfrak{B}^{\star}[N,0;\varphi(N)] <0$, otherwise, it has increasing nature. Furthermore, as we know, the coordinates of fixed points may be used to determine exact cosmological solutions at the fixed points themselves. One may utilize this method to analyze them like the taken procedure at the previous example.\\
However, it must be noted that one can perform all steps (from the definition of normalized variables to end) in usual ways, and finally he may use the $\mathfrak{B}\text{-function}$ method to analyze the obtained results. Both these approaches would be equivalent; the analysis is the important part, not the approach. We only wanted to indicate this feature as well.

In higher order theories of gravity, if one adopts the $\mathfrak{B}\text{-function}$ to consider the systems as autonomous systems, then he is back to a thing which is equivalent to the approach suggested by S. Carloni \cite{Carloni}. In higher order theories of gravity, we must utilize the connectional formulas presented at the end of the section (\ref{The A.R.M.}).\\

As already noticed, it is not our objective to carry a complete discussion out, hence, we terminate this section here and the further proceeding may easily be done by the reader as the studied example is a well-known case which has been investigated in several papers.\\

\noindent\hrmybox{}{\section{Conclusion}}\vspace{5mm}

In order to the analysis of the reconstruction methods, phase space, and exact solutions of the alternative theories of gravity, a new approach, $\mathfrak{B}\text{-Function}$ Method, was suggested. This new method imposes some conditions on parameters via a new function, $\mathfrak{B}\text{-Function}$, which is the Lie derivative of $\ln(\mathcal{F}(x))$ along a vector field, $\mathbf{X}=x+c$, on a singleton $Q=\{x\}$. This method renders the physically admissible domains of the parameters and helps to perform true data analysis. Based on this new perspective, at a very short time interval, all conditions are related to one of two classes. Each class was built on four new theorems. It may be stated that almost all the constant parameters are tightly coupled, hence one does not allow to select the arbitrary amounts of the constants in the data analysis in order to justify some of the important events such as phase crossing, late-time-accelerated expansion, and etcetera.

It was stated that our new approach may also be straightforwardly developed to formulate the analysis of the early inflation era based on the observational data, because the $\mathfrak{B}\text{-function}$ appears in the formulas of that era as well, for example, we indicated that the slow-roll parameters may also be expressed in terms of this function.

And finally, two examples were given in order to clarify the manner of using the method.\\

\section*{\noindent\goldmybox{red}{\vspace{3mm} Acknowledgments \vspace{3mm}}}

This work has been supported financially by Research Institute for Astronomy $\&$ Astrophysics of Maragha (RIAAM) under research project No. 1/6025-2.\\

\noindent\hrmybox{}{\section{Supplement: Proof of the $\mathfrak{B}\text{-Function}$ Method \label{Proofss}}}\vspace{5mm}

In this section, four new theorems required for the `Class-1' (based on the decreasing scalar field with time) are presented. The theorems of the second class may be performed in similar manners.\\

\begin{theo}{thm:pythagoras}\label{STARM1}
Let $a=a(t)$ be the scale factor of the universe. If with the reconstruction method, one has $a(\varphi)=\tilde{a}$, where $\tilde{a}$ is a real function of the scalar field, $\varphi=\varphi(t)$, then the following physical conditions
\begin{empheq}[box={\mymath[colback=blue!23,drop lifted shadow]}]{equation}\label{BBBT1}
\mathfrak{B}[t,0;a]>0,
\end{empheq}
\begin{empheq}[box={\mymath[colback=blue!23,drop lifted shadow]}]{equation}\label{BT}
\mathfrak{B}[\varphi,\beta_{0};\tilde{a}]<0,
\end{empheq}
where $\beta_{0}$ is an arbitrary constant, together with\\
\begin{empheq}[box={\mymath[colback=blue!20,drop lifted shadow]}]{equation}\label{BT120}
\mathfrak{B}[t,0;\varphi]<0,
\end{empheq}
which is equivalent to
\begin{empheq}[box={\mymath[colback=blue!20,drop lifted shadow]}]{equation}\label{BT120}
\mathfrak{B}[z,1;\grave{\varphi}]>0,
\end{empheq}
where $\grave{\varphi}=\varphi(z)$, must be established. In the anisotropic frameworks, these conditions must be held for each of the scale factors.\\
\end{theo}
\begin{prf}{thm:pythagoras}\label{SPARM1}
From the mathematical point of view, accelerating and also decelerating expansion pictures of the universe, provided by various astronomical and cosmological observations, imply that the scale factor must be strictly increasing not monotonically. On the other hand, the \textbf{absolute amount} of the scale factor (i.e. $|a(t)|$) is important, since its sign only shows the direction of expanding. Therefore, one must have
\begin{align}\label{BT2}
&\frac{\mathrm{d}|a|}{\mathrm{d}t}>0 \quad \longrightarrow \quad
\frac{\mathrm{d}|a|}{\mathrm{d}t}=\frac{a\dot{a}}{|a|}
\stackrel{\times \frac{a}{a}}{=\joinrel=\joinrel=\joinrel=}
\underbrace{\frac{a^2}{|a|}}_{>0}\frac{\dot{a}}{a}>0
\nonumber \\ & \longrightarrow \quad
t\frac{\dot{a}}{a}=tH_{a}=\mathfrak{B}[t,0;a]>0
\end{align}
or equivalently
\begin{align}\label{BT2.1}
&\frac{\mathrm{d}|a|}{\mathrm{d}t}>0 \nonumber \\ & \longrightarrow \frac{\mathrm{d}|\tilde{a}|}{\mathrm{d}t}
=\left(\frac{\mathrm{d}|\tilde{a}|}{\mathrm{d}\varphi}\right)
\left(\frac{\mathrm{d}\varphi}{\mathrm{d}t}\right)
=\left(\frac{\tilde{a} \frac{\mathrm{d}\tilde{a}}{\mathrm{d}\varphi}}{|\tilde{a}|}\right)\dot{\varphi}
\nonumber \\ &
\stackrel{\times \frac{(\varphi+\beta_{0}) \tilde{a}}{(\varphi+\beta_{0}) \tilde{a}}}{=\joinrel=\joinrel=\joinrel=\joinrel=\joinrel=}
\left(\frac{\varphi+\beta_{0}}{\tilde{a}}
\left(\frac{\mathrm{d}\tilde{a}}{\mathrm{d}\varphi} \right)\right) \underbrace{\left(\frac{{\tilde{a}}^2}{|\tilde{a}|}\right)}_{>0} \left(\frac{\dot{\varphi}}{\varphi+\beta_{0}}\right)>0.
\end{align}
Because the absolute amount of the scalar field of the class-1 must decrease with time (in general, monotonically decreasing), so we have
\begin{align}\label{BT3}
&\frac{\mathrm{d}|\varphi|}{\mathrm{d}t}\leq 0 \quad \longrightarrow \quad
\frac{\mathrm{d}|\varphi|}{\mathrm{d}t}=\frac{\varphi\dot{\varphi}}{|\varphi|}
\stackrel{\times \frac{\varphi}{\varphi}}{=\joinrel=\joinrel=\joinrel=}
\underbrace{\frac{\varphi^2}{|\varphi|}}_{>0} \frac{\dot{\varphi}}{\varphi}\leq 0
\nonumber \\ & \longrightarrow \quad
\frac{\dot{\varphi}}{\varphi}\leq 0.
\end{align}
On the other hand, one may redefine $\varphi:=\varphi+\beta_{0}$ which gives
\begin{align}\label{U001}
\frac{\dot{\varphi}}{\varphi}
=\frac{\dot{\overbrace{(\varphi+\beta_{0})}}}{\varphi+\beta_{0}}
=\frac{\dot{\varphi}}{\varphi+\beta_{0}} \leq 0,
\end{align}
but, it can here be only strictly decreasing, viz.,
\begin{align}\label{U004}
\frac{\dot{\varphi}}{\varphi} = \frac{\dot{\varphi}}{\varphi+\beta_{0}} <0
\end{align}
instead, because (\ref{BT2.1}) cannot be zero. Therefore, using (\ref{BT2.1}) and (\ref{U004}) we get two physical conditions:
\begin{align}\label{U002}
&\frac{\varphi+\beta_{0}}{\tilde{a}}
\left(\frac{\mathrm{d}\tilde{a}}{\mathrm{d}\varphi} \right)=\mathfrak{B}[\varphi,\beta_{0};\tilde{a}]<0,
\end{align}
\begin{align}\label{behzadsss}
t\frac{\dot{\varphi}}{\varphi}=\mathfrak{B}[t,0;\varphi]<0.
\end{align}
According to the connectional formula (\ref{BCF5}),
\begin{align*}
\mathfrak{B}[t,0;\mathcal{F}(t)]=-\mathfrak{B}[t,0;a(t)] \; \mathfrak{B}[z,1;\mathcal{F}(z)],
\end{align*}
where $\mathcal{F}$ is an arbitrary function, and eq. (\ref{BT2}), $\mathfrak{B}[t,0;a(t)]>0$, we easily conclude that
\begin{align*}
\mathfrak{B}[z,1;\mathcal{F}(z)] \propto -\mathfrak{B}[t,0;\mathcal{F}(t)].
\end{align*}
Therefore, taking $\mathcal{F}=\varphi$ and pursuant to (\ref{behzadsss}), the condition
\begin{align}
\mathfrak{B}[z,1;\varphi(z)] > 0
\end{align}
is obtained. $\Box$ \\

Unlike some papers in which the negative amounts of $m$ have been allowed, here, we find that it is not true, because one trivial consequence of (\ref{BT2}) is that $m>0$, because of $H_{1}=mH_{2}$. Furthermore, the observational data also confirms this.\\
\end{prf}

\begin{theo}{thm:pythagoras}\label{STARM2}
Let $\varphi=\varphi(t)$ be the nonzero scalar field. If we consider it as a function of the scale factor, $a=a(t)$, (i.e. $\tilde{\varphi}=\varphi(a)$) then the following condition must be satisfied
\begin{empheq}[box={\mymath[colback=cyan!30,drop lifted shadow]}]{equation}\label{BT3.3}
\mathfrak{B}[a,\alpha_{0};\tilde{\varphi}] < 0,
\end{empheq}
where $\alpha_{0}$ is an arbitrary constant.\\
\end{theo}
\begin{prf}{thm:pythagoras}\label{SPARM2}
According to the `Theorem \ref{STARM1}', one has $\mathfrak{B}[t,0;\varphi]<0$, therefore it gives
\begin{equation*}\label{BT0012}\begin{split}
&\frac{\dot{\varphi}}{\varphi}< 0 \\& \rightarrow \;
\frac{1}{\tilde{\varphi}}\frac{\mathrm{d}\tilde{\varphi}}{\mathrm{d}a}
\frac{\mathrm{d}a}{\mathrm{d}t}
\stackrel{\times \frac{a+\alpha_{0}}{a+\alpha_{0}}}
{=\joinrel=\joinrel=\joinrel=\joinrel=\joinrel=\joinrel=}
\frac{a+\alpha_{0}}{\tilde{\varphi}} \; \frac{\mathrm{d}\tilde{\varphi}}{\mathrm{d}a}
\underbrace{\frac{\dot{a}}{(a+\alpha_{0})}}_{>0} < 0.
\end{split}\end{equation*}
The term $\dot{a}/(a+\alpha_{0})$ is positive since according to (\ref{BT2}) and by the redefinition $a:=a+\alpha_{0}$ one has
\begin{align*}
\frac{\dot{a}}{a}=\frac{\dot{\overbrace{(a+\alpha_{0})}}}{a+\alpha_{0}}
=\frac{\dot{a}}{a+\alpha_{0}}>0.
\end{align*}
Therefore, we easily get\\
\begin{equation*}\label{787878}
\frac{a+\alpha_{0}}{\tilde{\varphi}}\; \frac{\mathrm{d}\tilde{\varphi}}{\mathrm{d}a}
=\mathfrak{B}[a,\alpha_{0};\tilde{\varphi}] < 0.
\end{equation*}
\text{ }
\end{prf}
\begin{theo}{thm:pythagoras}\label{STARM3}
Let $F=F(\varphi)$ be a nonzero scalar function which is dependent on the scalar field $\varphi$ (For example, $F$ can be a scalar field potential, a coupling function with curvature/torsion/electromagnetic field and etcetera). Every form of this function in each very short time interval must satisfy one of the following conditions:
\begin{empheq}[box={\mymath[colback=green!38,drop lifted shadow]}]{equation}\label{BT110}
\mathfrak{B}[\varphi,\phi_{0};F] \geq 0,
\end{empheq}
or
\begin{empheq}[box={\mymath[colback=pink!50,drop lifted shadow]}]{equation}\label{BT110.1}
\mathfrak{B}[\varphi,\phi_{0};F] \leq 0,
\end{empheq}
where $\phi_{0}$ is an arbitrary constant. The former one belongs to the class of the decreasing scalar functions with time,
\begin{empheq}[box={\mymath[colback=green!38,drop lifted shadow]}]{equation}\label{22BT110}
\mathfrak{B}[t,0;\mathbb{F}] \leq 0,
\end{empheq}
or equivalently
\begin{empheq}[box={\mymath[colback=green!38,drop lifted shadow]}]{equation}\label{1368122BT110}
\mathfrak{B}[z,1;\grave{F}] \geq 0,
\end{empheq}
where $\mathbb{F}=F(t)$ and $\grave{F}=F(z)$, and the later one belongs to the class of the increasing scalar functions with time,
\begin{empheq}[box={\mymath[colback=pink!50,drop lifted shadow]}]{equation}\label{22BT110.1}
\mathfrak{B}[t,0;\mathbb{F}] \geq 0,
\end{empheq}
or equivalently
\begin{empheq}[box={\mymath[colback=pink!50,drop lifted shadow]}]{equation}\label{1368222BT110.1}
\mathfrak{B}[z,1;\grave{F}] \leq 0.
\end{empheq}
Furthermore, if we consider all the functions as functions of the scale factor, $a=a(t)$, then for the function $F$ written in terms of the scale factor instead of the scalar field (i.e. $\tilde{F}=F(a)$, where $\tilde{F}$ is a real function of the scale factor) the conditions
\begin{empheq}[box={\mymath[colback=green!45,drop lifted shadow]}]{equation}\label{BT1}
\mathfrak{B}[a,\gamma_{0};\tilde{F}] \leq0,
\end{empheq}
\begin{empheq}[box={\mymath[colback=pink!45,drop lifted shadow]}]{equation}\label{BT1234}
\mathfrak{B}[a,\gamma_{0};\tilde{F}] \geq0,
\end{empheq}\\
for (\ref{BT110}) and (\ref{BT110.1}), must also be held respectively (i.e. In a case, the function $F$ must satisfy one of these set of conditions: 1. Eqs.(\ref{BT110}) and (\ref{BT1}); or, 2. Eqs. (\ref{BT110.1}) and (\ref{BT1234})). Here, $\gamma_{0}$ is an arbitrary constant. The equal signs correspond to the constant function, $F=f_{0}$. Note that in each set of the paired conditions, when one of the conditions is nonnegative, then one another is nonpositive.\\
\end{theo}
\text{  } \\

\begin{prf}{thm:pythagoras}\label{SPARM3}
A scalar function is monotonically increasing or decreasing versus its variable at each very short interval. Therefore, at each very short time interval, there are two types of the scalar functions assuming `Class-1 (D.S.F.)' for the scalar field: 1. The functions which the absolute amounts of them are of decreasing nature with respect to \textit{time}, such as $F\sim \varphi^2$, $F\sim \varphi^4$, and etcetera;  2. The functions which the absolute amounts of them are of increasing nature with respect to \textit{time}, such as $F\sim \exp(-\varphi),$ $F\sim \varphi^{-2},\cdots$. These mean that a scalar function can be monotonically decreasing or increasing versus time in throughout its evolution range. However, there are some forms of the function $F$ which are mixtures of two fundamental cases in throughout their evolution ranges, but, one can separate such cases into the monotonic parts. Furthermore, one may convert hybrid cases to monotonic cases in throughout of interval of interest by tuning constants of problem.\\
It is worth mentioning that in data analysis we can first consider the conditions, and then the treatments of the scalar function $F$ versus time may be obtained spontaneously.\\
Therefore, for a decreasing scalar function with time,
\begin{equation}\label{astagh}
\frac{\mathrm{d}|\mathbb{F}|}{\mathrm{d}t}
=\frac{\dot{\mathbb{F}}}{\mathbb{F}}\leq 0,
\end{equation}
which is equivalent to
\begin{equation}\label{astagh13}
t\frac{\dot{\mathbb{F}}}{\mathbb{F}}=\mathfrak{B}[t,0;\mathbb{F}]\leq 0,
\end{equation}
one has
\begin{equation}\label{BT1289}\begin{split}
&\frac{\mathrm{d}|\mathbb{F}|}{\mathrm{d}t}
=\frac{\mathrm{d}|F|}{\mathrm{d}\varphi}
\frac{\mathrm{d}\varphi}{\mathrm{d}t} \\&
\stackrel{\times \frac{(\varphi+\phi_{0} ) F}{(\varphi+\phi_{0}) F}}{=\joinrel=\joinrel=\joinrel=\joinrel=\joinrel=\joinrel=}
\left(\frac{(\varphi+\phi_{0} ) F^{\prime}}{F}\right)
\underbrace{\left(\frac{F^2}{|F|}\right)}_{>0}
\underbrace{\left(\frac{\dot{\varphi}}{\varphi+\phi_{0}}\right)}_{<0}\leq 0.
\end{split}\end{equation}
Pursuant to `Theorem \ref{STARM1}' and by the redefinition $\varphi:=\varphi+\phi_{0}$ yielding
\begin{equation*}
\frac{\dot{\varphi}}{\varphi}= \frac{\dot{\overbrace{(\varphi+\phi_{0})}}}{(\varphi+\phi_{0})} =\frac{\dot{\varphi}}{\varphi+\phi_{0}},
\end{equation*}
the third term in (\ref{BT1289}) would be negative. Therefore, we get
\begin{equation}\label{BT12}\begin{split}
\frac{(\varphi+\phi_{0})}{F}\left(\frac{\mathrm{d}F}{\mathrm{d}\varphi}\right) =\mathfrak{B}[\varphi,\phi_{0};F]\geq 0.
\end{split}\end{equation}
Note that, the constant, $\phi_{0}$, has been added in order to simplify the calculations, for example, in the potentials of the form $V=V_{0}(\varphi-\varphi_{0})^n$ it would be helpful, as by taking $\phi_{0}=-\varphi_{0}$, (\ref{BT12}) leads to $n$, while, if we do not put $\phi_{0}$ in (\ref{BT12}), then it yields $n\varphi / (\varphi-\varphi_{0})$.\\
Equivalently,
\begin{equation}\label{B13}\begin{split}
&\frac{\mathrm{d}|\mathbb{F}|}{\mathrm{d}t}\leq 0 \\ &\longrightarrow \frac{\mathrm{d}|\mathbb{F}|}{\mathrm{d}t}
=\left(\frac{\mathrm{d}|\tilde{F}|}{\mathrm{d}a}\right)
\left(\frac{\mathrm{d}a}{\mathrm{d}t}\right)
=\left(\frac{\tilde{F} \frac{\mathrm{d}\tilde{F}}{\mathrm{d}a}}{|\tilde{F}|}\right)\dot{a}\\ \\
&\stackrel{\times \frac{(a+\gamma_{0}) \tilde{F}}{(a+\gamma_{0}) \tilde{F}}}{=\joinrel=\joinrel=\joinrel=\joinrel=\joinrel=\joinrel=}
\left(\frac{a+\gamma_{0}}{\tilde{F}}\left(\frac{\mathrm{d}\tilde{F}}{\mathrm{d}a} \right)\right) \underbrace{\left(\frac{\tilde{F}^2}{|\tilde{F}|}\right)}_{>0} \underbrace{\left(\frac{\dot{a}}{a+\gamma_{0}}\right)}_{>0}\leq 0.
\end{split}\end{equation}
Therefore, one obtains
\begin{equation*}
\frac{(a+\gamma_{0})}{\tilde{F}}\left(\frac{\mathrm{d}\tilde{F}}{\mathrm{d}a} \right)=\mathfrak{B}[a,\gamma_{0};\tilde{F}]\leq 0,
\end{equation*}
where we have used the redefinition $a:=a+\gamma_{0}$ and thus
\begin{equation*}
\frac{\dot{a}}{a}=\frac{\dot{\overbrace{(a+\gamma_{0})}}}{a+\gamma_{0}}
=\frac{\dot{a}}{a+\gamma_{0}},
\end{equation*}
which according to `Theorem \ref{STARM1}' is a positive term.\\
According to the connectional formula (\ref{BCF5}),
\begin{align*}
\mathfrak{B}[t,0;\mathcal{F}(t)]&=-\mathfrak{B}[t,0;a(t)] \; \mathfrak{B}[z,1;\mathcal{F}(z)],
\end{align*}
where $\mathcal{F}$ is an arbitrary function, and eq. (\ref{BBBT1}), $\mathfrak{B}[t,0;a(t)]>0$, we easily conclude that
\begin{align}\label{adres}
\mathfrak{B}[z,1;\mathcal{F}(z)] \propto -\mathfrak{B}[t,0;\mathcal{F}(t)].
\end{align}
Therefore, taking $\mathcal{F}=F$ in (\ref{adres}) and using (\ref{astagh13}), we arrive at
\begin{align}
\mathfrak{B}[z,1;F(z)] \geq 0.
\end{align}

For the latter one (increasing scalar function), the proving way is the same, with the difference that in this case, one must start with $\frac{\mathrm{d}|\mathbb{F}|}{\mathrm{d}t}\geq 0$.\\
\end{prf}
\begin{theo}{thm:pythagoras}\label{STARM4}
Let $H$, $\breve{H}$, $\grave{H}$, and $\tilde{H}$ be the Hubble parameter written in terms of the time $t$, scale factor $a$, the redshift $z$, and the scalar field $\varphi$, respectively. Then, the expansion of the universe is accelerating if
\begin{empheq}[box={\mymath[colback=red!28,drop lifted shadow]}]{equation}\label{MUhy1}
\mathfrak{B}[t,0;H]>-\mathfrak{B}[t,0;a(t)],
\end{empheq}

\begin{empheq}[box={\mymath[colback=red!28,drop lifted shadow]}]{equation}\label{hy1}
\mathfrak{B}[a,0;\breve{H}]>-1,
\end{empheq}

\begin{empheq}[box={\mymath[colback=red!28,drop lifted shadow]}]{equation}\label{00hy1}
\mathfrak{B}[z,1;\grave{H}]<+1,
\end{empheq}

\begin{empheq}[box={\mymath[colback=red!28,drop lifted shadow]}]{equation}\label{000hy1}
\mathfrak{B}[\varphi,0;\tilde{H}]<\frac{-1}{\mathfrak{B}[a,0,\tilde{\varphi}]},
\end{empheq}
and is decelerating when
\begin{empheq}[box={\mymath[colback=red!28,drop lifted shadow]}]{equation}\label{MUhy2}
\mathfrak{B}[t,0;H]<-\mathfrak{B}[t,0;a(t)],
\end{empheq}

\begin{empheq}[box={\mymath[colback=red!28,drop lifted shadow]}]{equation}\label{hy2}
\mathfrak{B}[a,0;\breve{H}]<-1,
\end{empheq}

\begin{empheq}[box={\mymath[colback=red!28,drop lifted shadow]}]{equation}\label{00hy2}
\mathfrak{B}[z,1;\grave{H}]>+1,
\end{empheq}

\begin{empheq}[box={\mymath[colback=red!28,drop lifted shadow]}]{equation}\label{000hy2}
\mathfrak{B}[\varphi,0;\tilde{H}]>\frac{-1}{\mathfrak{B}[a,0,\tilde{\varphi}]}.
\end{empheq}
The status of the expansion of the universe is super-accelerating when the right hand side of the inequalities (\ref{MUhy1})-(\ref{000hy1}) is zero.\\
And the cases
\begin{empheq}[box={\mymath[colback=red!28,drop lifted shadow]}]{equation}\label{MUhy3}
\mathfrak{B}[t,0;H]= -\mathfrak{B}[t,0;a(t)],
\end{empheq}

\begin{empheq}[box={\mymath[colback=red!28,drop lifted shadow]}]{equation}\label{hy3}
\mathfrak{B}[a,0;\breve{H}]= -1,
\end{empheq}

\begin{empheq}[box={\mymath[colback=red!28,drop lifted shadow]}]{equation}\label{00hy3}
\mathfrak{B}[z,1;\grave{H}]= +1,
\end{empheq}

\begin{empheq}[box={\mymath[colback=red!28,drop lifted shadow]}]{equation}\label{000hy3}
\mathfrak{B}[\varphi,0;\tilde{H}]= \frac{-1}{\mathfrak{B}[a,0,\tilde{\varphi}]},
\end{empheq}
render the inflection point, namely shifting from decelerated to accelerated expansion.\\
Moreover, the universe is in phantom-like regime when
\begin{empheq}[box={\mymath[colback=red!28,drop lifted shadow]}]{equation}\label{MU11hy3}
\mathfrak{B}[t,0;H]>0,
\end{empheq}

\begin{empheq}[box={\mymath[colback=red!28,drop lifted shadow]}]{equation}\label{11hy3}
\mathfrak{B}[a,0;\breve{H}]>0,
\end{empheq}

\begin{empheq}[box={\mymath[colback=red!28,drop lifted shadow]}]{equation}\label{10011hy3}
\mathfrak{B}[z,1;\grave{H}]<0,
\end{empheq}

\begin{empheq}[box={\mymath[colback=red!28,drop lifted shadow]}]{equation}\label{100011hy3}
\mathfrak{B}[\varphi,0;\tilde{H}]<0,
\end{empheq}
and is in (quintessence/non-phantom)-like regime if
\begin{empheq}[box={\mymath[colback=red!28,drop lifted shadow]}]{equation}\label{111hy3}
\mathfrak{B}[t,0;H]<0,
\end{empheq}

\begin{empheq}[box={\mymath[colback=red!28,drop lifted shadow]}]{equation}\label{111hy3}
\mathfrak{B}[a,0;\breve{H}]<0,
\end{empheq}

\begin{empheq}[box={\mymath[colback=red!28,drop lifted shadow]}]{equation}\label{100111hy3}
\mathfrak{B}[z,1;\grave{H}]>0,
\end{empheq}

\begin{empheq}[box={\mymath[colback=red!28,drop lifted shadow]}]{equation}\label{1000111hy3}
\mathfrak{B}[\varphi,0;\tilde{H}]>0.
\end{empheq}
Consequently, the phase transition point is given by
\begin{empheq}[box={\mymath[colback=red!28,drop lifted shadow]}]{equation}\label{MU1111hy3}
\mathfrak{B}[t,0;H]= 0,
\end{empheq}

\begin{empheq}[box={\mymath[colback=red!28,drop lifted shadow]}]{equation}\label{1111hy3}
\mathfrak{B}[a,0;\breve{H}]= 0,
\end{empheq}

\begin{empheq}[box={\mymath[colback=red!28,drop lifted shadow]}]{equation}\label{1001111hy3}
\mathfrak{B}[z,1;\grave{H}]= 0,
\end{empheq}

\begin{empheq}[box={\mymath[colback=red!28,drop lifted shadow]}]{equation}\label{10001111hy3}
\mathfrak{B}[\varphi,0;\tilde{H}]= 0.
\end{empheq}
In the anisotropic backgrounds for each of the Hubble parameters, the conditions must be considered separately.\\
\end{theo}
\begin{prf}{thm:pythagoras}\label{SPARM4}
According to the definition of the deceleration parameter,
\begin{equation}\label{theorem4-1}
q=\frac{\mathrm{d}}{\mathrm{d}t}\left(\frac{1}{H} \right)-1,
\end{equation}
when $q<0$ and $q>0$ then the expansion of the universe are of accelerating and decelerating natures, respectively. Assuming $\breve{H}=H(a)$, (\ref{theorem4-1}) may be written as
\begin{align}\label{theorem4-2}
q &=\frac{-(\dot{H}+H^2)}{H^2}=\frac{-(a\breve{H}\breve{H}_{,a}
+\breve{H}^2)}{\breve{H}^2} \nonumber\\ &= - \mathfrak{B}[a,0;H(a)]-1.
\end{align}
Manifestly, pursuant to (\ref{theorem4-2}), or by taking $y=a$, $\mathcal{F}=H$, $c=0$, and $u=t$ in eq. (\ref{Chain rule}), one has:
\begin{align}
\mathfrak{B}[t,0;H]=\mathfrak{B}[t,0;a(t)] \; \mathfrak{B}[a,0;\breve{H}].
\end{align}
Therefore, the conditions may clearly be translated as follows
\begin{align}\label{theorem4-3}
q<0 &\Longleftrightarrow \left\{
\begin{array}{l}
\mathfrak{B}[t,0;H]>-\mathfrak{B}[t,0;a(t)]; \\
\mathfrak{B}[a,0;\breve{H}]>-1; \\
\mathfrak{B}[z,1;\grave{H}]<+1; \\
\mathfrak{B}[\varphi,0;\tilde{H}]< -\mathfrak{B}^{-1}[a,0;\tilde{\varphi}],
\end{array}\right.
\end{align}
\begin{align}
q>0 &\Longleftrightarrow \left\{
\begin{array}{l}
\mathfrak{B}[t,0;H]<-\mathfrak{B}[t,0;a(t)]; \\
\mathfrak{B}[a,0;\breve{H}]<-1; \\
\mathfrak{B}[z,1;\grave{H}]>+1; \\
\mathfrak{B}[\varphi,0;\tilde{H}]> -\mathfrak{B}^{-1}[a,0;\tilde{\varphi}],
\end{array}\right.
\end{align}
\begin{align}
q=0 &\Longleftrightarrow \left\{
\begin{array}{l}
\mathfrak{B}[t,0;H]=-\mathfrak{B}[t,0;a(t)]; \\
\mathfrak{B}[a,0;\breve{H}]=-1; \\
\mathfrak{B}[z,1;\grave{H}]=+1; \\
\mathfrak{B}[\varphi,0;\tilde{H}]= -\mathfrak{B}^{-1}[a,0;\tilde{\varphi}],
\end{array}\right.
\end{align}
where we have used the connectional formulas (\ref{BCF7}), viz.,
\begin{align}
\mathfrak{B}[a,0;H(a)]= - \mathfrak{B}[z,1;H(z)],
\end{align}
and (\ref{Chain rule}) in which we have set $y=\varphi$, $\mathcal{F}=H$, $c=0$, and $u=a$:
\begin{align}
\mathfrak{B}[a,0;H(a)]= \mathfrak{B}[a,0;\varphi(a)] \;  \mathfrak{B}[\varphi,0;H(\varphi)],
\end{align}
and also eq. (\ref{BT3.3}) in which we have taken $\alpha_{0}=0$. In the case $\dot{H}>0$, the type of the expansion of the universe is super-acceleration, hence its related conditions may easily be understood.\\
According to the relation between the effective EoS parameter, $\omega_{\mathrm{eff.}}$, and deceleration parameter $q$, viz.,
\begin{equation}\label{EOSQQQ}
\omega_{\mathrm{eff.}}=\frac{2q-1}{3},
\end{equation}
we consequently obtain the following relations:
\begin{align}\label{11111EOSQQQ}
&\text{(Quintessence/Non-Phantom)-Like Regime:} \nonumber\\
&\omega_{\mathrm{eff.}}>-1 \; \Longleftrightarrow  \left\{
\begin{array}{l}
\mathfrak{B}[t,0;H]<0; \\
\mathfrak{B}[a,0;\breve{H}]<0; \\
\mathfrak{B}[z,1;\grave{H}]>0; \\
\mathfrak{B}[\varphi,0;\tilde{H}]>0,
\end{array}\right.
\end{align}
\begin{align}
&\text{Phantom-Like Regime:}\nonumber\\
&\omega_{\mathrm{eff.}}<-1 \; \Longleftrightarrow
\left\{
\begin{array}{l}
\mathfrak{B}[t,0;H]>0;\\
\mathfrak{B}[a,0;\breve{H}]>0;\\
\mathfrak{B}[z,1;\grave{H}]<0; \\
\mathfrak{B}[\varphi,0;\tilde{H}]<0,
\end{array}\right.
\end{align}

\begin{align}
&\text{Phantom Divide Line (Phase Transition Point):}\nonumber\\
&\omega_{\mathrm{eff.}}=-1  \; \Longleftrightarrow
\left\{
\begin{array}{l}
\mathfrak{B}[t,0;H]=0; \\
\mathfrak{B}[a,0;\breve{H}]=0; \\
\mathfrak{B}[z,1;\grave{H}]=0; \\
\mathfrak{B}[\varphi,0;\tilde{H}]=0.
\end{array}\right.
\end{align}
\text{ }
\end{prf}
\begin{tcolorbox}[colback=white,
colframe=green]
Pursuant to (\ref{BCF5}) and eq. (\ref{BBBT1}), the special case $\mathcal{F}(t)=a(t)$ leads to $\mathfrak{B}[z,1,a(z)]=-1$ which is the closed form of the well-known relation $z+1=(a_{0}/a)$. Because this relation cannot reveals the expansion of the universe, hence we did not write the $\mathfrak{B}\text{-function}$ for $a(z)$ in Tables (\ref{RMA}) and (\ref{RMA2}).
\end{tcolorbox}
\text{ }\\
\text{ }
\hrule \hrule \hrule \hrule \hrule \hrule

\end{document}